\documentclass[aps,
prb,
twocolumn,
superscriptaddress,
longbibliography,
floatfix,
amsfonts,
amssymb,
amsmath]{revtex4-2}

\usepackage[final]{pdfpages}

\makeatletter
\AtBeginDocument{\let\LS@rot\@undefined}
\makeatother

\usepackage{bm,latexsym,mathrsfs,enumerate,bbm}
\usepackage{upgreek}
\usepackage[table,x11names]{xcolor}
\usepackage[breaklinks=true,
unicode=true,
urlcolor = RoyalBlue4,
colorlinks = true,
citecolor = RoyalBlue4,
linkcolor = blue
]{hyperref}
\usepackage{graphicx}
\graphicspath{{figs/}}
\usepackage{lmodern}
\usepackage{mathtools}
\usepackage{pifont}
\usepackage{multirow}
\usepackage[cal=boondoxo,frak=boondox,scr=rsfs]{mathalfa}
\renewcommand{\vec}[1]{\bm{#1}}

\DeclareMathOperator{\sech}{sech}

\begin{document}
\title{Geometric tailoring of domain wall traps and diodes in ferromagnetic nanostripes}

\author{Dmytro Karakuts}
\email{karakutsd@knu.ua}
\affiliation{Taras Shevchenko National University of Kyiv, 01601 Kyiv, Ukraine}

\author{Kostiantyn V. Yershov}
\email{k.yershov@ifw-dresden.de}
\affiliation{Leibniz-Institut f\"ur Festk\"orper- und Werkstoffforschung, IFW Dresden, 01171 Dresden, Germany}
\affiliation{Bogolyubov Institute for Theoretical Physics of the National Academy of Sciences of Ukraine, 03143 Kyiv, Ukraine}

\author{Denis D. Sheka}
\email{sheka@knu.ua}
\affiliation{Taras Shevchenko National University of Kyiv, 01601 Kyiv, Ukraine}

\date{October 25, 2026}


%
%

\begin{abstract}
	
Geometric tailoring of magnetic responses is a central theme in modern magnetism. By employing cross-sectional geometric modulation, we investigate how geometric variations influence the dynamics of field-driven domain walls. This approach enables the realization of a domain-wall trap, achieved through a nanostripe constriction, and a domain-wall diode, realized via a wedge-shaped expansion. We demonstrate that domain walls can be efficiently pinned by cross-sectional modulation, with the depinning field exhibiting strong dependence on the cross-sectional gradient. Under weak applied fields, the domain wall undergoes localized oscillations near the pinning site; under strong fields, this pinning breaks down, reproducing the typical diode characteristics. Theoretical calculations based on a biaxial ferromagnet model and a collective-variable approach are in excellent agreement with micromagnetic simulations. Finally, we delineate the parameter regime for realistic samples in which these predicted effects are most pronounced.
	
\end{abstract}

\maketitle


\section{Introduction}
The interplay between geometry and material properties has become a cornerstone of modern condensed matter physics, driving new developments in nanomagnetism with numeral applications~\cite{Gubbiotti25,Makarov22a,Vazquez20}. A particularly active research direction in modern magnetism is the study of topological magnetic textures,~\cite{Zang18} such as domain walls (DWs) and skyrmions, which are leading candidates for next-generation nonvolatile, high-density, and energy-efficient data storage and logic devices~\cite{Vermeulen24,Zhang25a}.

Recent studies have demonstrated that the geometry of magnetic nanostructures can be harnessed to manipulate various spin textures. In particular, nanostripes and nanowires exhibit a range of geometry-induced phenomena, including DW automotion, a promising mechanism for the low-power transport of magnetic information. While DW automotion in relatively wide (quasi-2D) nanostripes is typically accompanied by notable deformation of the DW profile~\cite{Chauleau10,Nikonov14,Richter16,Mawass17}, narrow nanostripes operate in a quasi-1D regime. Two primary mechanisms drive geometry-governed DW automotion in this 1D limit: first, curvature-gradient driving originating from an exchange-induced emergent Dzyaloshinskii--Moriya interaction~\cite{Yershov18a}; and second, spatial variations in the nanostripe cross section~\cite{Yershov23,Karakuts24b}. These mechanisms offer energy-efficient routes for DW manipulation that require no external stimuli, such as magnetic fields or electric currents. Furthermore, the ability to control DW behavior at the nanoscale through engineered geometry renders such structures ideal information carriers for advanced spintronic applications~\cite{Skoric22}.

Interest in these systems is further amplified by rapid advances in nanoscale fabrication techniques, including focused electron-beam-induced deposition~\cite{Huth18,Fernandez-Pacheco20}, two-photon lithography~\cite{Hunt20}, and multiphoton lithography~\cite{Farinha25}. Notably, recent fabrication efforts have confirmed geometrically induced DW automotion and demonstrated that variations in the nanostripe cross section significantly influence DW motion, sometimes dominating over curvature gradients~\cite{Skoric22,Fernandez-Roldan19,Berganza25}.

DW motion can be controlled by modifying the nanowire cross section in several ways, including cutting notches, fabricating multisegment wires, or gradually varying the cross-sectional profile. By introducing notches or protrusions, one can engineer pinning sites or potential barriers to control DW motion, as demonstrated in multiple experimental studies~\cite{Klaui08,Petit08,Petit10}. Nanowire notches can also enhance the DW propagation velocity~\cite{Yuan15}. The strength of DW pinning at notches has been shown to depend on chirality and can even induce spontaneous DW depinning in the absence of an external field~\cite{Garcia-Sanchez11}. When the notches are asymmetric, they may function as DW diodes~\cite{Himeno05,Whyte15}, enabling the construction of DW ratchets via a sawtooth profile~\cite{Franken12a} and offering an alternative to designs based on alternating material properties~\cite{Bran18,Berganza17}.

Fabricating nanowires with segments of different cross-sectional areas provides a means of tailoring the DW potential landscape, with wider sections acting as energy barriers. Alternating wider and narrower segments can serve as DW traps~\cite{Rodriguez16,Nasirpouri19}. Bamboo-like nanowires, featuring a series of expanded segments along their length, have been demonstrated to enable precise field-controlled manipulation of pinned DWs~\cite{Berganza16,Bran21}. DW pinning in such cross-section-modulated structures has also been investigated theoretically and numerically~\cite{Riz20,Fernandez-Roldan19}. Owing to the magnetic stability of each segment and the ability to write independent magnetic orientations in the narrower regions, these structures have been proposed as candidate information-storage media~\cite{Saez21}. Cross-sectional modulation of multisegment nanowires has furthermore been explored as a route to geometrically tuned DW nucleation and propagation~\cite{Fradet25,Arzuza17}.

Recently, a theoretical framework has been developed to describe the magnetic configurations of nanostripes with a spatially varying cross-sectional area~\cite{Yershov23}. Using this framework, the effects of cross-sectional variations on DW dynamics in ferromagnetic nanostripes were analyzed, and it was shown that the cross-sectional gradient can act as a source of DW pinning~\cite{Yershov23,Karakuts24b}. However, none of these studies has considered the role of an external magnetic field, which is essential for both experimental verification and practical device applications.

Here, we aim to theoretically investigate the interplay between an applied magnetic field and cross-sectional modulation in governing DW dynamics in ferromagnetic nanostripes. We illustrate these effects through two representative geometries: a straight nanostripe with a localized constriction and a wedge-shaped nanostripe.


\section{Model}
We consider a thin, narrow ferromagnetic nanostripe with a rectangular cross section whose width and thickness are sufficiently small that the magnetization remains uniform across the entire cross section, while the stripe length is much larger than its transverse dimensions. In this regime, the normalized magnetization $\vec{m}(x,t)=\vec{M}/M_s$ depends only on time $t$ and the longitudinal coordinate $x$, with $M_s$ denoting the saturation magnetization. The nanostripe geometry is defined by the domain
$x\in(-\infty,\infty)$, $y\in\bigl[-w(x)/2,\,w(x)/2\bigr]$, $z\in\bigl[-d(x)/2,\,d(x)/2\bigr]$, where $w(x)$ and $d(x)$ are the local width and thickness, respectively, with $w(x)\geq d(x)$. For convenience we introduce dimensionless coordinates $\xi=x/\ell$ and a dimensionless cross-sectional area $S(\xi)=\mathcal{S}(\xi\ell)/\ell^{2}$, where $\mathcal{S}(x)$ is the cross-sectional area and $\ell=\sqrt{A/(4\pi M_s^{2})}$ is the exchange length, with $A$ being the exchange stiffness constant.

The magnetic properties of such a nanostripe are described by a classical biaxial-ferromagnet model. The energy density of the nanostripe in an external field, normalized by $\mathcal{E}_0=4\pi M_s^{2}\ell^{3}$, comprises three contributions,
$\mathcal{E}=\mathcal{E}^{\textsc{x}}+\mathcal{E}^{\textsc{a}}+\mathcal{E}^{\textsc{z}}$.
The first term,
$\mathcal{E}^{\textsc{x}}=S(\xi)\,m'_{\nu}\,m'_{\nu}$,
corresponds to the exchange interaction, where a prime denotes differentiation with respect to $\xi$.
The second term,
$\mathcal{E}^{\textsc{a}}=S(\xi)\bigl[k_{\textsc{a}}(1-m_{x}^{2})+k_{\textsc{p}}\,m_{z}^{2}\bigr]$,
is the anisotropy contribution, with effective coefficients
$k_{\textsc{a}}=K_{\textsc{a}}/(4\pi M_s^{2})+k_{\textsc{a}}^{\textsc{ms}}$ and
$k_{\textsc{p}}=K_{\textsc{p}}/(4\pi M_s^{2})+k_{\textsc{p}}^{\textsc{ms}}$,
characterizing easy-axis and easy-plane anisotropies, respectively. Here $K_{\textsc{a}}>0$ and $K_{\textsc{p}}>0$ are the magnetocrystalline anisotropy constants, while $k_{\textsc{a}}^{\textsc{ms}}$ and $k_{\textsc{p}}^{\textsc{ms}}$ account for the magnetostatic contribution~\cite{Aharoni98}. The last term, $\mathcal{E}^{\textsc{z}}=-S(\xi)\,\vec{h}\cdot\vec{m}$, is the Zeeman energy density arising from an external field $\vec{H}$, where $\vec{h}=\vec{H}/(4\pi M_s)$ is the normalized field. In the present work we consider a static, uniform field directed along the stripe length, $\vec{h}=h\,\hat{\bm{x}}$.
Throughout, we impose the unit-magnitude constraint $|\vec{m}|=1$ and parameterize the order parameter as
$\vec{m}=\hat{\bm{x}}\cos\theta+\hat{\bm{y}}\sin\theta\cos\phi+\hat{\bm{z}}\sin\theta\sin\phi$.

To analyze both the static and dynamic properties of DWs, we employ a collective-variable approach~\cite{Malozemoff79,Slonczewski72} based on the Slonczewski $q$--$\Phi$ ansatz~\cite{Slonczewski72,Slonczewski73,Slonczewski79}:
\begin{equation}\label{eq:qPhiAnsatz}
	\cos\theta(\xi,\bar{t})=-\tanh\!\frac{\xi-q(\bar{t})}{\Delta(\bar{t})},
	\qquad
	\phi(\xi,\bar{t})=\Phi(\bar{t}),
\end{equation}
where $q$ and $\Phi$ are conjugate collective variables representing the DW position and phase, respectively; $\Delta$ is the DW width; and $\bar{t}=\omega_{0}t$ is a dimensionless time, with $\omega_{0}=4\pi\gamma_{0}M_s$ setting the characteristic frequency of the system and $\gamma_{0}$ being the gyromagnetic ratio.

Although this framework is general and applicable to a wide range of geometries, we focus here on two representative configurations: a nanostripe with a localized constriction and a wedge-shaped nanostripe. In both cases we employ the narrow-DW approximation $\Delta\ll l_{s}$, where $l_{s}\sim S/S'$ is the length scale of the cross-sectional variation.

\begin{figure}[t]
	\includegraphics[width=\columnwidth]{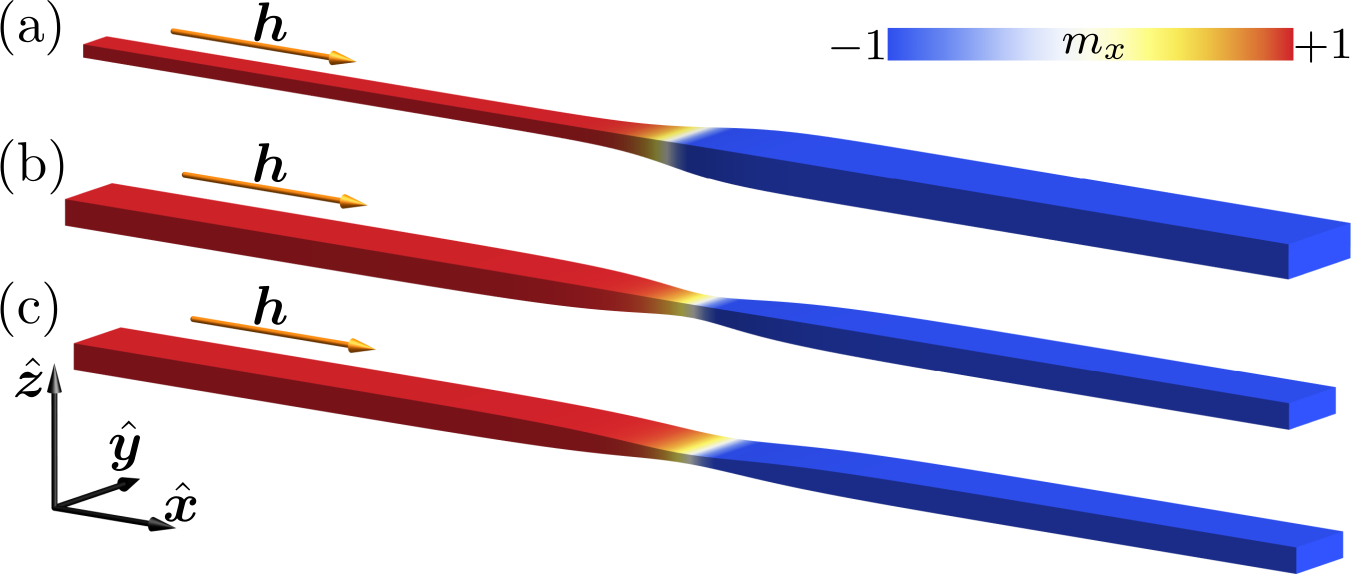}
	\caption{
		Schematic illustration of DWs in ferromagnetic nanostripes of different geometries: (a)~wedge-shaped nanostripe with a constant aspect ratio, (b)~nanostripe with a localized constriction and a constant aspect ratio, and (c)~nanostripe with a localized constriction produced by thickness variation alone. The cross-sectional profiles are defined by Eqs.~\eqref{eq:WS_CSfunction},~\eqref{eq:LC_CSfunction}, and~\eqref{eq:VARK_CSfunction}, respectively. The color scheme indicates the local magnetization direction, and the orange arrow denotes the applied-field direction.
		\label{fig:GeoIllustration}%
	}
\end{figure}


\section{Domain wall in a nanostripe with a constant aspect ratio}
\label{sec:CA}

\begin{figure*}
	\centering
	\includegraphics[width=\textwidth]{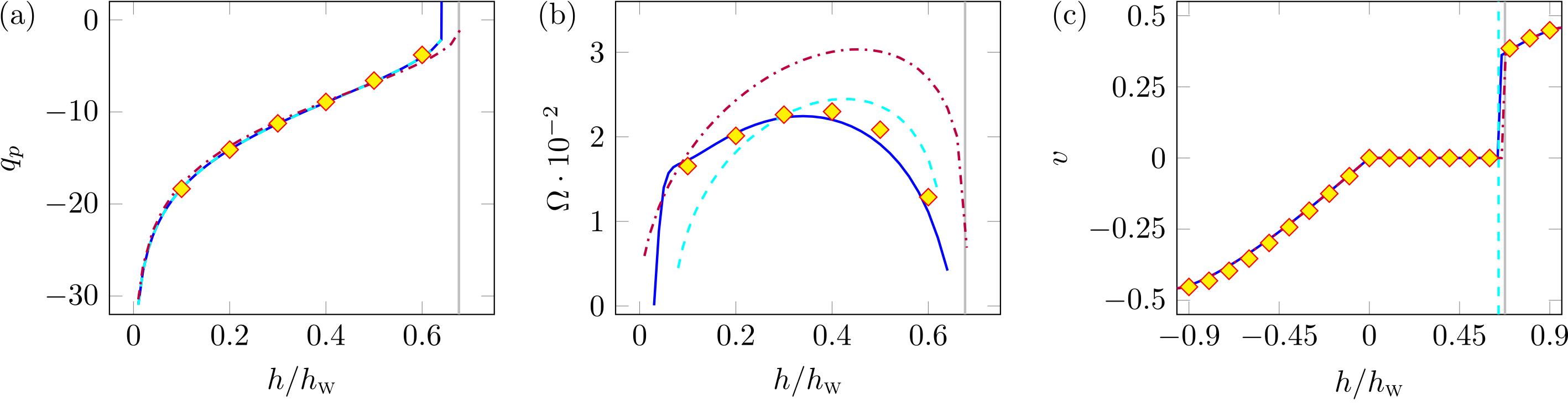}
	\caption{Dynamics and pinning of DW in a wedge-shaped nanostripe: (a)~DW pinning-position offset induced by an external magnetic field; (b)~frequency of DW oscillations near the pinning position as a function of the applied field; and (c)~DW velocity in the proposed DW diode as a function of the applied field. The dash-dotted red lines represent predictions from the narrow-DW approximation based on Eqs.~\eqref{eq:CA_pinning_condition}, \eqref{eq:CA_Frequencies}, and the numerical solution of Eq.~\eqref{eq:CA_EOM}, plotted in panels (a), (b), and (c), respectively. The vertical gray line marks the depinning field defined by Eq.~\eqref{eq:CA_depinning_field}. The solid blue curve shows the numerical solution of the equations of motion for a non-narrow DW; see Eq.~(S25) in the Supplemental Material~\cite{Note1}. The dashed cyan lines correspond to the analytical predictions for a non-narrow DW given by Eqs.~(S26), (S30), and (S27), plotted in panels (a), (b), and (c), respectively. Symbols represent the results of full-scale micromagnetic simulations using NMAG~\cite{Fischbacher07}; see the Supplemental Material for details~\cite{Note1}. The nanostripe geometry is defined by Eq.~\eqref{eq:WS_CSfunction}, utilizing the material parameters of Permalloy~\cite{Note2} and a damping coefficient of $\alpha=0.07$. The applied-field amplitude is measured in units of the Walker field, $h_{\textsc{w}}=\alpha k_{\textsc{p}}$.}
	\label{fig:WScombined}%
\end{figure*}
\footnotetext[2]{For simulations, the material parameters of Permalloy are used: the exchange stiffness $A=2.6\times10^{-6}~\text{erg/cm}~(26~\text{pJ/m})$ and the saturation magnetization $M_{s}=800~\text{G}~(800~\text{kA/m})$.}

We investigate the effects of a variable cross section on field-induced DW dynamics. In this configuration, both the stripe width $w(x)$ and thickness $d(x)$ vary along the stripe length. We begin by considering the case in which the cross section maintains a constant aspect ratio, $\delta=w(x)/d(x)=\text{const}$. This condition directly affects the anisotropy coefficients: specifically, the magnetostatic contribution to the effective anisotropy is determined by the aspect ratio, as detailed in Sec.~SM-1 of the Supplemental Material (SM)~\footnote[1]{See Supplemental Material \texttt{link provided by the publisher} for details of the analytical calculations and micromagnetic simulations, including Refs.~\cite{Yershov23,Karakuts24b,Aharoni98,Thiaville06,Porter04,Fischbacher07}.}. Consequently, for a constant aspect ratio, the anisotropy parameters $k_{\textsc{a}}$ and $k_{\textsc{p}}$ remain uniform throughout the system.

Here, we consider nanostripes whose cross-sectional area is defined as
\begin{equation}\label{eq:S_definition}
	S(\xi)=S_{0}\bigl[1+s(\xi)\bigr], \quad s(\xi)>-1, \quad s'(\pm\infty)=0,
\end{equation}
as illustrated in Figs.~\ref{fig:GeoIllustration}(a) and~\ref{fig:GeoIllustration}(b).

By applying the ansatz \eqref{eq:qPhiAnsatz} with the cross-sectional profile \eqref{eq:S_definition} to the DW energy and integrating over $\xi$, we obtain the effective DW energy in a nanostripe with a constant aspect ratio (up to an additive constant):
\begin{equation}\label{eq:CA_Edw}
	E^{\textsc{dw}}=2S(q)\left[\frac{1}{\Delta}+\Delta\bigl(k_{\textsc{a}}+k_{\textsc{p}}\sin^{2}\Phi\bigr)\right]-2h\,\mathcal{I}(q),
\end{equation}
where $\mathcal{I}(\xi)=\int_{-\infty}^{\xi}S(x)\,\mathrm{d}x$. This description of the magnetic energy remains valid provided the cross-sectional variations are sufficiently smooth~\cite{Mawass17,Skoric22,Petit10}.

Using this effective energy, we derive the equations of DW motion in a nanostripe with a constant aspect ratio by applying the Lagrange--Rayleigh formalism; see Sec.~SM-2 of the SM~\cite{Note1} for details:
\begin{equation}\label{eq:CA_EOM}
	\begin{split}
		\alpha\,\frac{\dot{q}}{\Delta}+\dot{\Phi} &= \mathscr{F}(q),\\
		\frac{\dot{q}}{\Delta}-\alpha\,\dot{\Phi} &= k_{\textsc{p}}\sin 2\Phi,
	\end{split}
\end{equation}
where an overdot denotes the dimensionless time derivative and $\mathscr{F}(q)$ is the effective driving field,
\begin{equation}\label{eq:CA_effective_field}
	\mathscr{F}(q) = h -\frac{2}{\Delta}\frac{\partial_{q}s(q)}{1+s(q)}.
\end{equation}
In Eqs.~\eqref{eq:CA_EOM} and~\eqref{eq:CA_effective_field}, the DW width is treated as a slave variable, $\Delta(\bar{t})\equiv\Delta\bigl[\Phi(\bar{t})\bigr]=\bigl(k_{\textsc{a}}+k_{\textsc{p}}\sin^{2}\Phi\bigr)^{-1/2}$~\cite{Thiaville04,Thiaville06}, since the damping parameter is typically very small, $\alpha\ll1$, and the characteristic relaxation time is proportional to $\alpha$~\cite{Mougin07}.

\begin{figure*}[t]
	\centering
	\includegraphics[width=\textwidth]{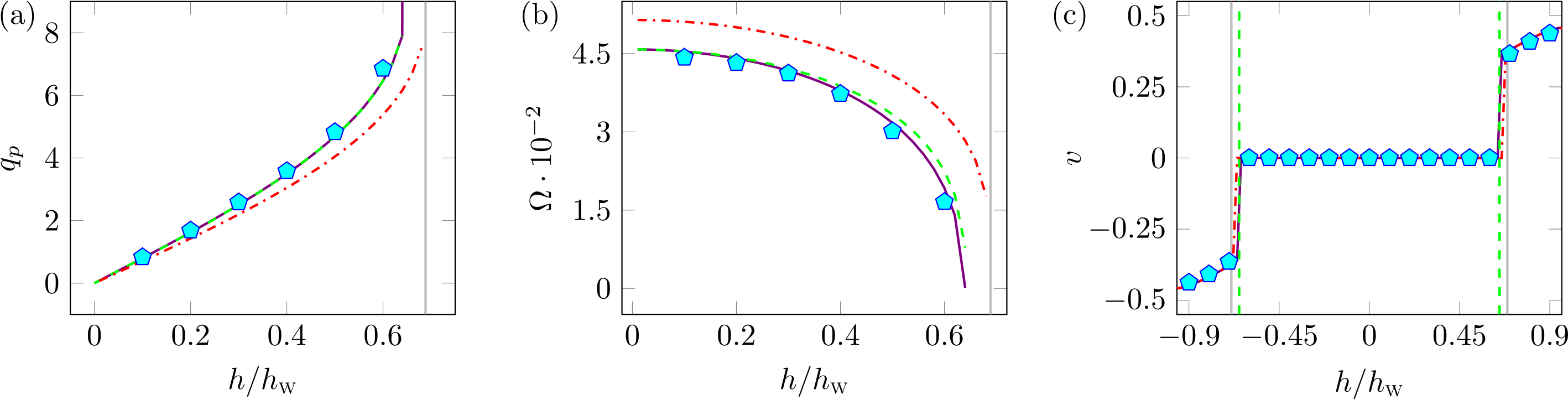}
	\caption{Dynamics and pinning of DW in a constricted nanostripe: (a)~DW pinning-position offset induced by an external magnetic field; (b)~frequency of DW oscillations near the pinning position as a function of the applied field; and (c)~DW velocity in the proposed DW trap as a function of the applied field. The dash-dotted red lines represent predictions from the narrow-DW approximation based on Eqs.~\eqref{eq:CA_pinning_condition}, \eqref{eq:CA_Frequencies}, and the numerical solution of Eq.~\eqref{eq:CA_EOM}, plotted in panels (a), (b), and (c), respectively. The vertical gray line marks the depinning field defined by Eq.~\eqref{eq:CA_depinning_field}. The solid purple curve shows the numerical solution of the equations of motion for a non-narrow DW; see Eq.~(S25) in the Supplemental Material~\cite{Note1}. The dashed green lines correspond to the analytical predictions for a non-narrow DW given by Eqs.~(S26), (S30), and (S27), plotted in panels (a), (b), and (c), respectively. Symbols represent the results of full-scale micromagnetic simulations using NMAG~\cite{Fischbacher07}; see the Supplemental Material for details~\cite{Note1}. The nanostripe geometry is defined by Eq.~\eqref{eq:LC_CSfunction}, utilizing the material parameters of Permalloy~\cite{Note2} and a damping coefficient of $\alpha=0.07$. The applied-field amplitude is measured in units of the Walker field, $h_{\textsc{w}}=\alpha k_{\textsc{p}}$.}
	\label{fig:LCcombined}%
\end{figure*}

The equations of motion \eqref{eq:CA_EOM} are structurally similar to those for an undeformed stripe~\cite{Thiaville06}, differing only in the effective driving field \eqref{eq:CA_effective_field}. The key distinction is the additional term in $\mathscr{F}(q)$ arising from the cross-sectional gradient. As evident from \eqref{eq:CA_EOM}, the DW dynamics are governed by the competition between the applied magnetic field and the cross-sectional gradient. For sufficiently small fields, the DW remains pinned at the cross-sectional variation; larger fields drive it through the deformation. Note that DW pinning by large curvature gradients is already well established~\cite{Yershov15b,Sheka15}.

\subsection{Domain wall pinning and depinning processes}

For a pinned DW, its position and phase are constant, so we set $\dot{q}=0$ and $\dot{\Phi}=0$. Substituting these conditions into \eqref{eq:CA_EOM} yields the pinning condition for the DW:
\begin{equation}\label{eq:CA_pinning_condition}
	h -\frac{2}{\Delta_{0}}\frac{\partial_{q}s(q_{p})}{1+s(q_{p})}=0, \qquad \sin\Phi_{p}=0,
\end{equation}
where $q_{p}$ and $\Phi_{p}$ are the DW pinning position and phase, respectively. For a pinned DW, the width can be treated as a constant, $\Delta_{0}=1/\sqrt{k_{\textsc{a}}}$. The resulting field dependence of the DW pinning position is shown in Figs.~\ref{fig:WScombined}(a) and~\ref{fig:LCcombined}(a).

As evident from \eqref{eq:CA_pinning_condition}, if the applied field is sufficiently small, the DW shifts to a new pinning position where the field is balanced by the cross-sectional gradient. The DW pinning position is thus the smallest solution of $\mathscr{F}(q_{p})=0$ that is greater than or equal to the initial DW position $q_{0}$. Conversely, if the applied field is large enough that the first term in \eqref{eq:CA_pinning_condition} exceeds the second term everywhere along the nanostripe, the DW depins and escapes the inhomogeneous region. The external field required for DW depinning, obtained from \eqref{eq:CA_pinning_condition}, is
\begin{equation}\label{eq:CA_depinning_field}
	h_{d}=\frac{2}{\Delta_{0}}\max_{q\in[q_{0},+\infty)}\left[\frac{\partial_{q}s(q)}{1+s(q)}\right].
\end{equation}
A comparison of the depinning field \eqref{eq:CA_depinning_field} with the results of micromagnetic simulations is shown in Figs.~\ref{fig:WScombined} and~\ref{fig:LCcombined} by the vertical gray lines.

\subsection{Domain wall linear dynamics}

For fields $h<h_{d}$, the DW does not settle immediately into its pinned position; instead, damped oscillations around the pinning position are expected. To study these oscillations, we linearize the equations of DW motion \eqref{eq:CA_EOM} in the vicinity of the pinning position, which yields
\begin{equation}\label{eq:CA_lind_EOM_VF}
	\begin{bmatrix}
		\dot{\widetilde{q}}\\[2pt]
		\dot{\widetilde{\Phi}}
	\end{bmatrix}
	\approx
	\begin{bmatrix}
		-\alpha\Omega_{g} & \Delta_{0}\Omega_{\textsc{b}}\\[2pt]
		-\Omega_{g}/\Delta_{0} & -\alpha\,\Omega_{\textsc{b}}
	\end{bmatrix}
	\begin{bmatrix}
		\widetilde{q}\\[2pt]
		\widetilde{\Phi}
	\end{bmatrix}.
\end{equation}
Solving the linear system in Eq.~\eqref{eq:CA_lind_EOM_VF} yields damped DW oscillations, $\widetilde{q}\propto\sin(\Omega\bar{t})\,e^{-\eta\bar{t}}$ and $\widetilde{\Phi}\propto\cos(\Omega\bar{t})\,e^{-\eta\bar{t}}$, with frequency $\Omega$ and damping rate $\eta$ given by
\begin{equation}\label{eq:CA_Frequencies}
	\begin{aligned}
		&\Omega\approx\sqrt{\Omega_{\textsc{b}}\Omega_{g}}, \qquad \eta=\frac{\alpha\left(\Omega_{g}+\Omega_{\textsc{b}}\right)}{2},\\
		&\Omega_{g}=\Omega_{s}-\Omega_{h}, \qquad
		\Omega_{s}=\frac{2}{1+\alpha^{2}}\frac{\partial^{2}_{q}s(q_{p})}{1+s(q_{p})},\\
		&\Omega_{\textsc{b}}=\frac{2}{1+\alpha^{2}}\,k_{\textsc{p}}, \qquad
		\Omega_{h}=\frac{\Delta_{0}^{2}}{2\left(1+\alpha^{2}\right)}h^{2}.
	\end{aligned}
\end{equation}

The DW oscillation frequency \eqref{eq:CA_Frequencies} reduces to the well-known DW eigenfrequency for a localized constriction in a straight nanostripe at zero field~\cite{Yershov23}. Indeed, the anisotropy term $\Omega_{\textsc{b}}$ and the cross-sectional-gradient term $\Omega_{s}$ are identical to their zero-field counterparts. However, the application of an external field introduces a novel correction $\Omega_{h}\propto h^{2}$ in \eqref{eq:CA_Frequencies}. The resulting field dependence of the DW eigenfrequency is shown in Figs.~\ref{fig:WScombined}(b) and~\ref{fig:LCcombined}(b).

\subsection{Domain wall inertial mass}

By considering the equation of DW motion \eqref{eq:CA_EOM} in the vicinity of the pinning position $q_{p}$ in the absence of damping ($\alpha\to0$) and applying a small-angle expansion $\Phi=\Phi_{p}+\varphi$, one finds that the DW velocity is proportional to the phase deviation, $v=2k_{\textsc{p}}\Delta_{0}\varphi$. Substituting this relation into the effective DW energy \eqref{eq:CA_Edw} gives
\begin{equation}\label{eq:CA_Edw_using_v}
	E^{\textsc{dw}}=E_{\text{const}}+\frac{1}{2}S_{0}\bigl[1+s(q_{p})\bigr]\frac{v^{2}}{k_{\textsc{p}}\Delta_{0}}.
\end{equation}
The DW energy thus comprises two contributions: a constant term and a term proportional to $v^{2}$, analogous to the standard expression for kinetic energy. This implies that the DW possesses an effective inertial mass, first identified by D\"{o}ring for DW motion in a nanostripe with a uniform cross section~\cite{Thiaville06,Doring48}. Cross-sectional variations modify this D\"{o}ring mass as follows:
\begin{equation}\label{eq:CA_Doring_mass}
	m_{\textsc{d}}=\bigl[1+s(q_{p})\bigr]\mu_{\textsc{d}},
\end{equation}
where $\mu_{\textsc{d}}=S_{0}/(k_{\textsc{p}}\Delta_{0})$ is the D\"{o}ring mass of a nanostripe with a constant cross section~\cite{Thiaville06,Doring48}. The DW inertial mass therefore scales with the local cross-sectional area, being larger in wider sections than in thinner ones.


\section{Domain-wall diode and trap}
\label{sec:diodeandtrap}

We now consider two geometrically distinct configurations of practical interest: a wedge-shaped nanostripe and a nanostripe with a localized constriction, illustrated in Figs.~\ref{fig:GeoIllustration}(a) and~\ref{fig:GeoIllustration}(b), respectively.

\subsection{Wedge-shaped nanostripes}

A wedge-shaped nanostripe is described by the cross-sectional-area profile \eqref{eq:S_definition} under the additional condition that the cross-sectional areas at the two ends differ; without loss of generality, we assume $s(-\infty) < s(+\infty)$. In this geometry, the cross-sectional gradient opposes DW motion in the positive $x$ direction when the DW originates in the narrow region. For fields below the depinning threshold, this results in DW blockage [Fig.~\ref{fig:WScombined}(a)], whereas for fields exceeding the depinning field \eqref{eq:CA_depinning_field}, the DW traverses the expansion [Fig.~\ref{fig:WScombined}(c)]. Crucially, this blocking is unidirectional: if the DW originates in the wide region and the field is applied in the opposite direction, it moves without hindrance. The wedge-shaped expansion therefore functions as a DW diode [Fig.~\ref{fig:WScombined}(c)], selectively permitting or blocking DW motion depending on the propagation direction.

The DW oscillation frequency near the pinning position is governed by Eq.~\eqref{eq:CA_Frequencies}. The frequency is primarily determined by the competition between the cross-sectional-gradient term $\Omega_{s}$ and the field-induced term $\Omega_{h}$. At small fields, both contributions are small, resulting in a low frequency. As the field increases, $\Omega_{s}$ grows faster than $\Omega_{h}$, causing the frequency to rise. However, as the field approaches the depinning field, $\Omega_{h}$ dominates and the frequency drops sharply [Fig.~\ref{fig:WScombined}(b)]. It is also worth noting that the DW oscillation frequency is enhanced by the easy-plane anisotropy.

\begin{figure*}[t]
	\centering
	\includegraphics[width=\textwidth]{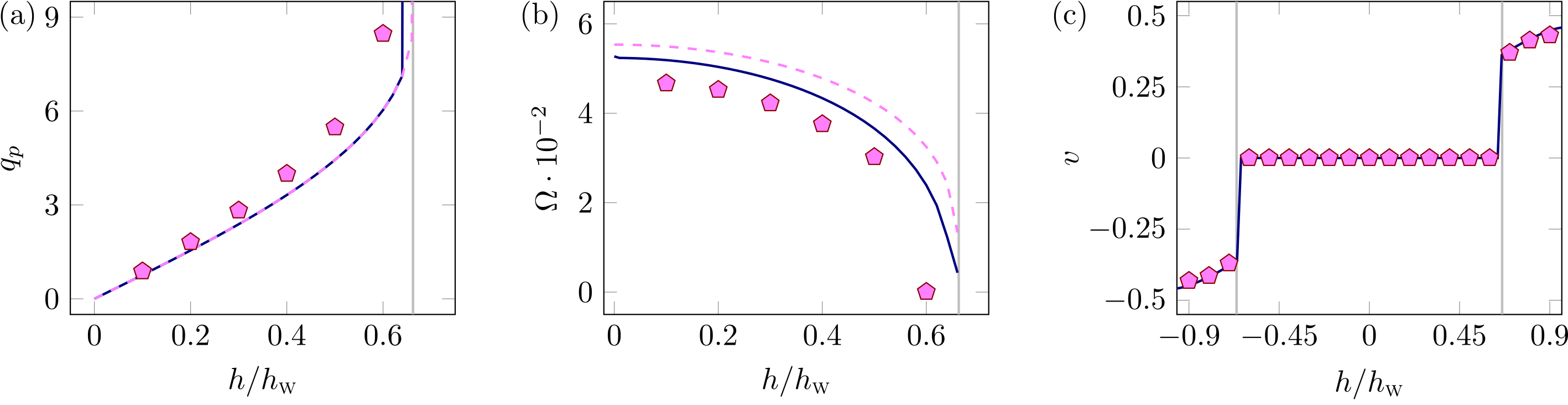}
	\caption{Dynamics and pinning of a domain wall in a constricted nanostripe with varying thickness: (a)~DW pinning-position offset induced by an external magnetic field; (b)~frequency of DW oscillations near the pinning position as a function of the applied field; and (c)~DW velocity in the proposed DW trap created by thickness variation as a function of the applied field. The solid blue curve shows the numerical solution of the equations of motion for the DW, Eq.~\eqref{eq:VARK_EOM}, while the dashed pink lines correspond to the analytical predictions given by Eqs.~\eqref{eq:VARK_pinning_condition} and~\eqref{eq:VARK_Frequencies}, plotted in panels (a) and (b), respectively. Symbols represent the results of full-scale micromagnetic simulations using NMAG~\cite{Fischbacher07}; see the SM for details~\cite{Note1}. The vertical gray line marks the depinning field defined by Eq.~\eqref{eq:VARK_depinning_field}, with the applied field normalized by $h_{\textsc{w}}=\alpha k_{\textsc{p}}(+\infty)$. The nanostripe geometry is defined by Eq.~\eqref{eq:VARK_CSfunction}, utilizing the material parameters of Permalloy~\cite{Note2} and a damping coefficient of $\alpha=0.07$.}
	\label{fig:VARKcombined}%
\end{figure*}

For a wedge-shaped nanostripe, the DW inertial mass in the wide section exceeds that in the narrow section. When the DW crosses the wedge-shaped expansion in the direction of increasing width, its inertial mass increases, giving rise to an effective resistance to motion.

To validate our theoretical predictions, we performed full-scale micromagnetic simulations of the Landau--Lifshitz--Gilbert equations using the NMAG code~\cite{Fischbacher07} (Fig.~\ref{fig:WScombined}). As a specific example, we considered a wedge-shaped nanostripe with a cross-sectional area described by
\begin{equation}\label{eq:WS_CSfunction}
	\mathcal{S}(x)=\mathcal{S}_{0}\left(1+\mathcal{s}\,\tanh\!\frac{x}{\mathtt{w}}\right), \qquad 0<\mathcal{s}<1,
\end{equation}
where $\mathcal{S}_{0}=5~\text{nm}^{2}$ is the reference cross-sectional area, $\mathcal{s}=0.15$ is the expansion magnitude, and $\mathtt{w}=60~\text{nm}$ is the characteristic expansion length; see the SM~\cite{Note1} for modeling details. The nanostripe aspect ratio is $\delta=3$.

Consistent with our model, the DW motion is strictly unidirectional, confirming the diode behavior of the wedge-shaped nanostripe. Motion in the positive direction is blocked, with the DW pinned at the wedge-shaped expansion for fields below the depinning threshold, whereas motion in the opposite direction proceeds unimpeded, with the DW rapidly accelerating to its maximum velocity [Fig.~\ref{fig:WScombined}(c)].

\subsection{Nanostripes with a localized constriction}

A second practically significant geometry is a nanostripe with a localized constriction [Fig.~\ref{fig:GeoIllustration}(b)], described by the cross-sectional-area profile \eqref{eq:S_definition} subject to the additional constraints $-1<s(\xi)\leq0$ and $s(\pm\infty)=0$.

Analogous to the wedge-shaped case, the DW is pinned for small applied fields, coming to rest at the position where the cross-sectional gradient balances the field [Fig.~\ref{fig:LCcombined}(a)]. For larger fields, depinning occurs [Fig.~\ref{fig:LCcombined}(c)] at the critical field given by \eqref{eq:CA_depinning_field}.

In contrast to the wedge-shaped geometry, the DW barrier associated with a localized constriction is bidirectional: if the DW originates within the constriction, its motion is impeded by the cross-sectional gradient irrespective of the field direction. Only when the applied field exceeds the depinning threshold can the DW cross the barrier and escape the constriction. This behavior identifies a localized constriction as a DW trap: the DW remains stationary at low fields and is released at high fields.

The process of DW settling into its pinned position again involves damped oscillations with a frequency determined by \eqref{eq:CA_Frequencies}. However, the oscillation frequency differs markedly from the wedge-shaped case [Fig.~\ref{fig:LCcombined}(b)], because the second-derivative term $\Omega_{s}$ is large near the pinning position, yielding high frequencies even at low fields. For a DW in a nanostripe with a localized constriction, increasing the field monotonically decreases the frequency, without the initial rise observed in the wedge-shaped geometry.

As in the wedge-shaped case, the reduction in cross-sectional area at the center of the constriction lowers the DW inertial mass relative to the constriction edges. When the DW exits the constriction, the cross-sectional area increases, causing a corresponding increase in the DW inertial mass that manifests as an effective resistance to motion.

Full-scale micromagnetic simulations were again performed to confirm these predictions (Fig.~\ref{fig:LCcombined}). As a specific example, we considered a nanostripe with a cross-sectional area described by
\begin{equation}\label{eq:LC_CSfunction}
	\mathcal{S}(x)=\mathcal{S}_{0}\left(1-\mathcal{s}\,\sech\!\frac{x}{\mathtt{w}}\right), \qquad 0<\mathcal{s}<1,
\end{equation}
with the following parameters: aspect ratio $\delta=3$, reference cross-sectional area $\mathcal{S}_{0}=5~\text{nm}^{2}$, constriction depth $\mathcal{s}=0.25$, and characteristic constriction length $\mathtt{w}=60~\text{nm}$.

The NMAG simulations are in good agreement with the model predictions for a localized constriction acting as a DW trap. For applied fields below the depinning threshold, the DW remains trapped within the constriction, with its pinning position determined by \eqref{eq:CA_pinning_condition}. Application of larger fields enables the DW to escape the constriction and subsequently accelerate rapidly to its maximum velocity [Fig.~\ref{fig:LCcombined}(c)].


\section{Domain wall in a nanostripe with a varying aspect ratio}
\label{sec:VARK}

Although a constant aspect ratio simplifies the theoretical description, experimentally it is often more convenient to fabricate nanostripes by varying only the thickness or width, resulting in a spatially varying aspect ratio. In this case, the anisotropy parameters become position-dependent, $k_{\textsc{a}}(\xi)$ and $k_{\textsc{p}}(\xi)$. The narrow-DW approximation remains applicable provided that $\Delta\ll l_{s}+l_{a}$ and $\Delta\ll l_{s}+l_{p}$, where $l_{a}\sim k_{\textsc{a}}/k_{\textsc{a}}'$ and $l_{p}\sim k_{\textsc{p}}/k_{\textsc{p}}'$.

Following the same procedure as in Sec.~\ref{sec:CA}, we obtain the effective DW energy
\begin{equation}\label{eq:VARK_Edw}
	E^{\textsc{dw}}=2S(q)\left\{\frac{1}{\Delta}+\Delta\left[k_{\textsc{a}}(q)+k_{\textsc{p}}(q)\sin^{2}\Phi\right]\right\}-2h\,\mathcal{I}(q),
\end{equation}
which is identical to \eqref{eq:CA_Edw} except that the anisotropy coefficients now depend on the DW position. For brevity, we abbreviate $k_{\textsc{a}}(q)$ and $k_{\textsc{p}}(q)$ as $k_{\textsc{a}}$ and $k_{\textsc{p}}$ throughout this section.

Applying the Lagrange--Rayleigh formalism to the effective energy yields a set of coupled equations of motion~\cite{Note1}:
 \begin{equation}\label{eq:VARK_full_EOM}
	\begin{aligned}
		\alpha\frac{\dot{q}}{\Delta}\!+\!\dot{\Phi} &=\!h\!-\!\frac{2}{\Delta}\frac{\partial_q s(q)}{1+s(q)}\!-\!\Delta\left(\partial_q k_ \textsc{a}\!+\!\sin^2\Phi \partial_q k_ \textsc{p}\right),\\
		\frac{\dot{q}}{\Delta}-\alpha\dot{\Phi}&=k_ \textsc{p}\sin{2\Phi},\\
		\alpha c\dot{\Delta}&=\frac{1}{\Delta}-\Delta\left(k_\textsc{a}+k_\textsc{p}\sin^2\Phi\right).
	\end{aligned}
\end{equation}
where $c$ is a geometric factor defined in the SM~\cite{Note1}. As in Sec.~\ref{sec:CA}, we treat the DW width as a slave variable,
$\Delta(\bar{t})\equiv\Delta\bigl[q(\bar{t}),\Phi(\bar{t})\bigr]=\bigl[k_{\textsc{a}}(q)+k_{\textsc{p}}(q)\sin^{2}\Phi\bigr]^{-1/2}$,
which is valid provided the DW-width relaxation is significantly faster than the DW motion. An estimate of the DW-width relaxation time (see the SM~\cite{Note1}) confirms that this condition holds for typical DW velocities $\dot{q}<1$ (corresponding to speeds below $\sim\!1$~km/s). The simplified equations of motion are then
\begin{subequations} \label{eq:VARK_EOM&F}
\begin{align}\label{eq:VARK_EOM}
&
\begin{aligned}
\alpha\,\frac{\dot{q}}{\Delta}+\dot{\Phi} &= \widetilde{\mathscr{F}}(q,\Phi),\\
\frac{\dot{q}}{\Delta}-\alpha\,\dot{\Phi} &= k_{\textsc{p}}\sin 2\Phi,
\end{aligned}
\intertext{where the effective driving field is}
\label{eq:VARK_effective_field}
&\widetilde{\mathscr{F}}(q,\Phi)=\mathscr{F}(q)-\Delta\left(\partial_{q}k_{\textsc{a}}+\sin^{2}\Phi\,\partial_{q}k_{\textsc{p}}\right).
\end{align}
\end{subequations}
The DW dynamics are thus governed by the competition among three contributions: the external magnetic field, the cross-sectional gradient, and the shape-anisotropy gradient.

As in the constant-aspect-ratio case, for small fields the DW shifts to a new pinning position where the field is balanced by the cross-sectional and anisotropy variations [Fig.~\ref{fig:VARKcombined}(a)]. The pinning condition now reads
\begin{equation}\label{eq:VARK_pinning_condition}
	\widetilde{\mathscr{F}}(q_{p},\Phi_{p})=0, \qquad \Phi_{p}\in\{0,\pi\}.
\end{equation}
Accordingly, the depinning field is modified by the shape-anisotropy gradient as well [Fig.~\ref{fig:VARKcombined}(c)]:
\begin{equation}\label{eq:VARK_depinning_field}
	h_{d}=\max_{q\in[q_{0},+\infty)}\left[\frac{2}{\Delta_{\star}}\frac{\partial_{q}s(q)}{1+s(q)}+\Delta_{\star}\,\partial_{q}k_{\textsc{a}}\right],
\end{equation}
where the local DW width at the pinning position is $\Delta_{\star}(q)=1/\sqrt{k_{\textsc{a}}(q)}$.

\begin{figure*}[t]
	\centering
	\includegraphics[width=\textwidth]{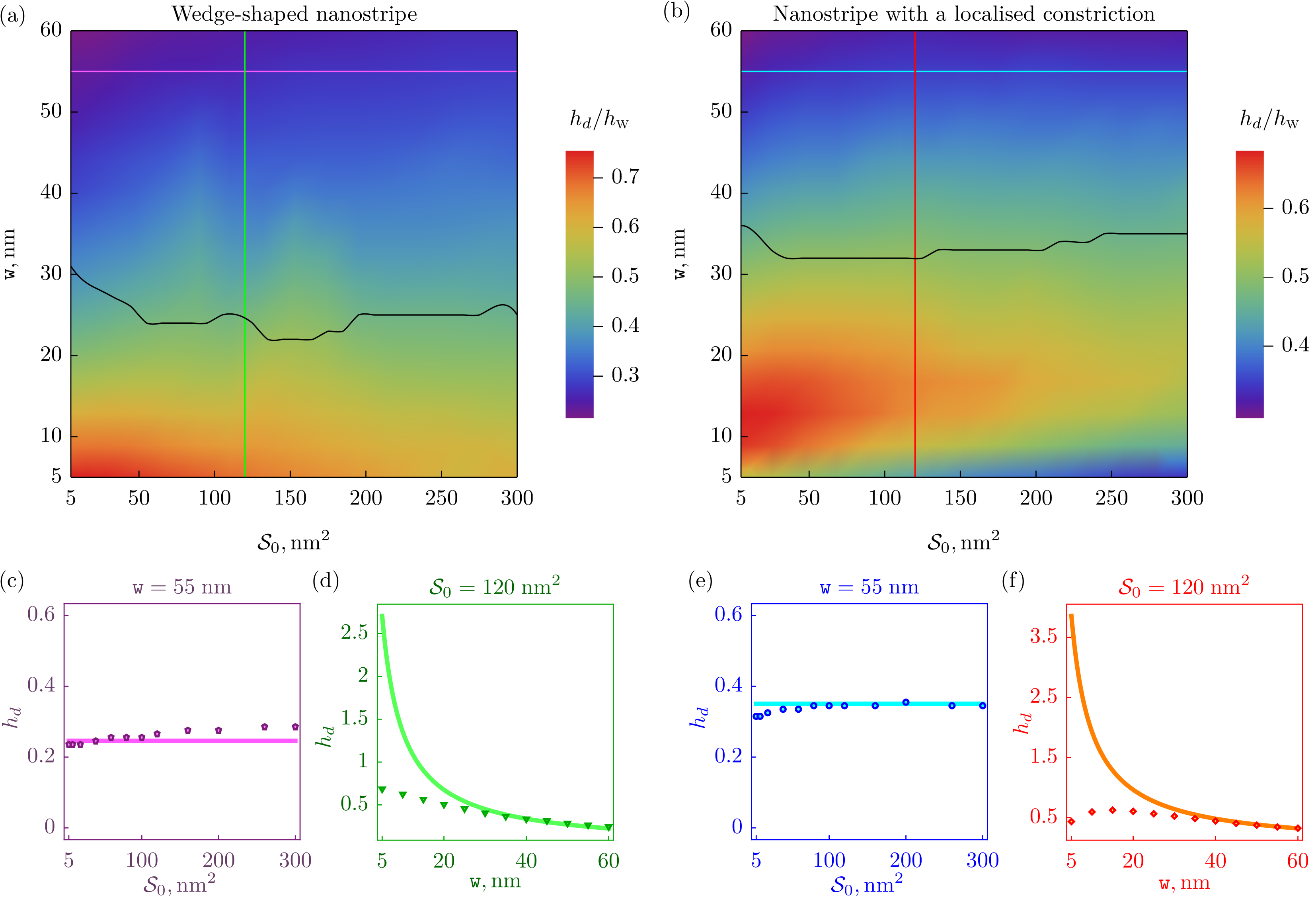}
	\caption{Parameter dependence of the domain-wall depinning field and validity regimes of the theoretical approximations. Panels (a) and (b) show two-dimensional parameter maps of the DW depinning field obtained from NMAG~\cite{Fischbacher07} micromagnetic simulations for a wedge-shaped nanostripe ($\mathcal{s}=0.05$, $\alpha=0.07$) and a nanostripe with a localized constriction ($\mathcal{s}=0.25$, $\alpha=0.15$), respectively. The color scale indicates the applied field required for DW depinning, normalized by $h_{\textsc{w}}=\alpha k_{\textsc{p}}$. The black isoline demarcates the range of validity of the narrow-DW approximation~\eqref{eq:CA_depinning_field}: above this contour, the narrow-DW theory accurately predicts the depinning field, whereas below it, the full non-narrow DW model is required. The colored vertical and horizontal lines indicate the locations of the corresponding one-dimensional cuts at fixed geometric parameters, which are presented in panels (c)--(f). Panels (c) and (d) display the cuts for the wedge-shaped case [panel (a)] at $w=55$~nm and $S_{0}=120$~nm$^{2}$, respectively. Panels (e) and (f) show the corresponding cuts for the nanostripe with a localized constriction [panel (b)]. In panels (c)--(f), symbols represent simulation results, while solid lines indicate theoretical predictions based on the narrow-DW approximation~\eqref{eq:CA_depinning_field}.}
	\label{fig:HdStability}%
\end{figure*}

The decaying oscillations exhibit a structure similar to that of the constant-aspect-ratio case [Fig.~\ref{fig:VARKcombined}(b)], with additional contributions arising in the DW oscillation frequency and damping rate:
\begin{equation}\label{eq:VARK_Frequencies}
	\begin{aligned}
		&\Omega_{g}=\Omega_{s}-\Omega_{h}+\Omega_{h\textsc{a}}-\Omega_{\textsc{a}},\\
		&\Omega_{h\textsc{a}}=\frac{3\Delta_{p}^{3}}{2\left(1+\alpha^{2}\right)}h\,\partial_{q}k_{\textsc{a}}(q_{p}),\\
		&\Omega_{\textsc{a}}=\frac{\Delta_{p}^{2}}{2\left(1+\alpha^{2}\right)}\left\{2\partial_{q}^{2}k_{\textsc{a}}(q_{p})-3\left[\Delta_{p}\,\partial_{q}k_{\textsc{a}}(q_{p})\right]^{2}\right\},
	\end{aligned}
\end{equation}
where $\Delta_{p}=\Delta_{\star}(q_{p})$ is the DW width at the pinning position.

For the case of a varying aspect ratio, the DW inertial mass is also described by \eqref{eq:CA_Doring_mass}, although its behavior differs substantially because $k_{\textsc{p}}$ varies with the pinning position:
\begin{equation}\label{eq:VARK_Doring_mass}
	m_{\textsc{d}}=\frac{S(q_{p})}{k_{\textsc{p}}(q_{p})\,\Delta_{p}}.
\end{equation}

Our analytical predictions were verified by full-scale micromagnetic simulations. The nanostripe geometry was described by
\begin{align}\label{eq:VARK_dfunction}
	d(x)&=d_{0}\left(1-\mathcal{s}\,\sech\!\frac{x}{\mathtt{w}}\right), \qquad 0<\mathcal{s}<1,\\
	\label{eq:VARK_CSfunction} %
	\mathcal{S}(x)&=w_{0}\,d(x), 
\end{align}
with an initial aspect ratio $\delta=3$, reference cross-sectional area $\mathcal{S}_{0}=5$~nm$^{2}$, constriction depth $\mathcal{s}=0.2$, and characteristic constriction length $\mathtt{w}=60$~nm; see the SM~\cite{Note1} for details.

\section{Model limits}
\label{sec:model_limits}

In this section we discuss the applicability of our model to nanostripes with larger cross sections. To determine the limits of the theoretical model in terms of system size, we performed a series of micromagnetic simulations using the NMAG code~\cite{Fischbacher07}; see the SM~\cite{Note1} for details.

The simulations were carried out for progressively larger initial cross-sectional areas, $S_{0}$, ranging from $5$ to $300$~nm$^{2}$, and for characteristic variation lengths $\mathtt{w}$ ranging from $5$ to $60$~nm. Both nanostripes with a localized constriction \eqref{eq:LC_CSfunction} and wedge-shaped nanostripes \eqref{eq:WS_CSfunction} with constant aspect ratios were considered. For each geometry, we computed the relative error between the predicted depinning field \eqref{eq:CA_depinning_field} and the field required for DW depinning in the simulations. The narrow-DW model is considered valid when this error remains small ($\epsilon<20\%$), as illustrated in Fig.~\ref{fig:HdStability}.

As shown in Fig.~\ref{fig:HdStability}, two key factors determine the applicability of the theoretical model:
\begin{itemize}
	\item[(i)] The characteristic length of the cross-sectional variation should be sufficiently large compared to the DW width. Although for very thin nanostripes the model remains accurate even for small variations, the results become inaccurate when the characteristic constriction length is comparable to or smaller than the DW width. This is expected, as one of the key assumptions, $\Delta\ll S/S'$, no longer holds. For such geometries, equations derived without the narrow-DW approximation (see the SM~\cite{Note1}) may yield significantly improved accuracy. It is also worth noting that in nanostripes with a very narrow constriction, the DW can skip over the constriction entirely, which appears in Fig.~\ref{fig:HdStability} as a region of low depinning field in the bottom-right corner.
	\item[(ii)] The characteristic cross-sectional area of the nanostripe has a significant impact on the model's accuracy. As seen in Fig.~\ref{fig:HdStability}, the model gradually loses accuracy with increasing initial stripe area $S_{0}$, with substantial divergence for stripes with areas exceeding $200$~nm$^{2}$. However, if the constriction length is significantly larger than the DW width, the depinning field remains stable for considerably larger stripes. In addition, the nanostripe geometry itself plays a role: the results for wedge-shaped nanostripes remain accurate for significantly larger systems than those for nanostripes with a localized constriction. This discrepancy arises from the deformation of the DW profile for larger cross sections. When the transverse dimensions exceed the exchange length, the DW profile begins to deform and the system can no longer be treated as one-dimensional. It is well established that with increasing cross-sectional dimensions, the DW initially develops an asymmetric profile, while vortex walls become energetically favored at even larger cross sections~\cite{Hillebrands06}.
\end{itemize}


\section{Discussion}

In summary, utilizing the recently developed micromagnetic framework for stripes with varying cross sections~\cite{Yershov23}, we analyze DW motion in straight ferromagnetic nanostripes with varying cross sections under an external magnetic field.
This study focuses on two specific geometries: nanostripes with a localized constriction and wedge-shaped nanostripes, although our approach is general and applicable to a wide range of geometries. By employing the collective variable approach, we demonstrate that cross-section variations induce a competition between the effects of the external magnetic field and the cross-section gradient on DW dynamics.The cross-section gradient generates an internal driving force that, under certain conditions, counteracts the influence of the external magnetic field on the DW. The balance between these effects results in the formation of a DW pinning position, the value of which is determined for both geometries.

The application of a strong magnetic field moves the DW out of the cross-section deformation.
The depinning field $h_d$, sufficient for the DW to escape the deformation in both geometries, is determined primarily by the ratio between the cross-section gradient and the area of the deformation.
It is noteworthy that even relatively small variations with large gradients, such as narrow but deep grooves, can significantly impact DW motion.
This effect can be exploited for the fabrication of DW diodes and traps by tailoring the nanostripe cross section.
A localized constriction acts as a DW trap by preventing DW motion at low magnetic fields and releasing the DW at larger fields.
Conversely, a wedge-shaped expansion operates as a DW diode by blocking DW motion in one direction while allowing free motion in the other.
This behavior is analogous to a DW diode based on functionally graded Dzyaloshinskii--Moriya interaction~\cite{Yershov20b}, where the coordinate dependence of the Dzyaloshinskii--Moriya parameter similarly leads to the filtering of DWs with specific topological charges and helicities.

In addition to the simple case of a constant width-to-thickness ratio, we consider the general case of varying shape anisotropy, including the experimentally significant case of solely varying thickness.
Varying the aspect ratio results in an additional component of the effective magnetic field corresponding to the gradient of the shape anisotropy coefficients.

Small deviations from the DW pinning position give rise to decaying oscillations.
By comparing the obtained oscillation frequencies, Eqs.~\eqref{eq:CA_Frequencies} and \eqref{eq:VARK_Frequencies}, with the DW eigenfrequencies in the absence of an external field~\cite{Yershov23}, we observe the following: (i)~Easy-plane anisotropy $k_\textsc{p}$ \textit{increases} the DW oscillation frequency for \textit{all} geometries. (ii)~For stripes with a constant aspect ratio, the presence of an external magnetic field introduces a new term proportional to the square of the field strength. (iii)~However, if the width-to-thickness ratio varies, two additional terms appear, corresponding to the gradient of shape anisotropy and its coupling with the external field.

Cross-section variations modify the inertial (D\"oring) DW mass~\cite{Doring48} as $m=\left[1+s(q)\right]\mu_\textsc{d}$, with nanostripe cross-section constrictions resulting in a reduction of the DW mass and expansions causing an increase.

Our results demonstrate that nanostripe cross-section tailoring is a viable strategy for DW manipulation.
The analytical results presented in this article have been verified by full-scale micromagnetic simulations.

In this work, we restricted our consideration to the influence of variable cross sections on the motion of the simplest topological excitations, DWs, in the most fundamental geometry, a planar stripe.
The interplay between spatially varying geometry and curvature-induced effects is expected to pave the way for novel applications in curvilinear spintronics and curvilinear magnonics~\cite{Sheka23}.
Furthermore, extending this approach to other magnetic topological excitations, such as skyrmions, promises to stimulate new avenues of research in skyrmionics~\cite{Koraltan26}.

\section*{Acknowledgments}

We appreciate discussions with A.~Fernández-Pacheco. We also thank Ulrike Nitzsche (IFW) for technical support. K.~Y. acknowledges the support of the Deutsche Forschungsgemeinschaft (DFG, German Research Foundation) through the grant No. YE 232/2-1. D.~S. acknowledges the support of the National Research Foundation of Ukraine via Grant No. 2025.07/0237 `Detection, analysis, and energy harvesting of microwave and terahertz signals in systems with spin diodes'. The simulations were made at the computing cluster of Taras Shevchenko National University of Kyiv \cite{unicc}. 

%
\clearpage 


\section*{Supplementary material}

# Supplemental Material for "Geometric tailoring of domain wall traps and diodes in ferromagnetic nanostripes"

Dmytro Karakuts,[1,*] Kostiantyn V. Yershov,[2,3,†] and Denis D. Sheka[1,‡]

[1] *Taras Shevchenko National University of Kyiv, 01601 Kyiv, Ukraine*
[2] *Leibniz-Institut für Festkörper- und Werkstoffforschung, IFW Dresden, 01171 Dresden, Germany*
[3] *Bogolyubov Institute for Theoretical Physics of the National Academy of Sciences of Ukraine, 03143 Kyiv, Ukraine*



This supplementary material provides detailed information on the analytical calculations and micromagnetic simulations of domain wall motion control in ferromagnetic nanostrips with varying cross sections under an external magnetic field.

## SM-1. MODEL OF A PLANAR FERROMAGNETIC NANOSTRIPE IN AN EXTERNAL MAGNETIC FIELD

Following the notations introduced in the main text, we describe a planar ferromagnetic stripe. Note that our approach is applicable to both stripes and wires.

For the case of a stripe, the nanostripe geometry can be defined as a spatial domain with the following bounds: $x \in (-\infty, \infty)$, $y \in [-w(x)/2, w(x)/2]$, $z \in [-d(x)/2, d(x)/2]$, where $w(x) = \sqrt{\delta(x)\mathcal{S}(x)}$ and $d(x) = \sqrt{\mathcal{S}(x)/\delta(x)}$ represent the nanostripe width and thickness, respectively. Here, $w(x) \geq d(x)$. The parameter $\mathcal{S}(x)$ denotes the cross-sectional area of the nanostripe, while $\delta(x) = w(x)/d(x)$ indicates the cross-sectional aspect ratio. For simplicity, we employ a normalized system by introducing dimensionless coordinates $\xi = x/\ell$, where $\ell = \sqrt{A/(4\pi M_s^2)}$ is the exchange length, with $A$ being the exchange constant.

It is important to note that the current micromagnetic framework requires the stripe thickness $d$ and width $w$ to be sufficiently small compared to the characteristic length scale of magnetic inhomogeneities $\ell_m$ (determined by material parameters such as exchange, anisotropy, and saturation magnetization) and the characteristic length scale of geometric inhomogeneities $\ell_g$ (determined by cross-sectional variations). Our analysis is limited by the following assumptions [1, 2]:

$$\frac{w^2}{\ell_g}, \frac{w^2}{\ell_m} \leq h \ll w \lesssim \ell_m, \ell_g \ll L, \tag{S1}$$

where $L$ is the length of the stripe.

Since the magnetic properties of such a stripe are described in the main text, we focus here on the derivation of the effective biaxial anisotropy term. As mentioned in the main text, the anisotropy energy density component is given by [1]:

$$\mathscr{E}^{\text{A}} = k_{\text{A}}\left(1 - m_x^2\right) + k_{\text{P}} m_z^2. \tag{S2}$$

The effective anisotropy parameters for the easy-axis and easy-plane types, $k_{\text{A}}$ and $k_{\text{P}}$, respectively, are defined as [1]:

$$k_{\text{A}} = \frac{K_{\text{A}}}{4\pi M_s^2} + k_{\text{A}}^{\text{MS}}, \qquad k_{\text{P}} = \frac{K_{\text{P}}}{4\pi M_s^2} + k_{\text{P}}^{\text{MS}}, \tag{S3}$$

where $K_{\text{A}} > 0$ and $K_{\text{P}} > 0$ are the magnetocrystalline anisotropy constants, and $k_{\text{A}}^{\text{MS}}$ and $k_{\text{P}}^{\text{MS}}$ arise from magnetostatic interactions [3].

For a uniformly magnetized stripe with a rectangular cross-section, the magnetostatic energy is known to be equivalent to an effective *shape* anisotropy with the following parameters [3–5]:

$$k_{\text{A}}^{\text{MS}} = \frac{1}{2\pi}\left[\frac{1-\delta^2}{2\delta}\ln\left(1+\delta^2\right) + \delta\ln\delta + 2\arctan\frac{1}{\delta}\right], \tag{S4}$$

$$k_{\text{P}}^{\text{MS}} = \frac{1}{2} - 2k_{\text{A}}^{\text{MS}}, \tag{S5}$$

where $\delta = w/d$ is the aspect ratio of the nanostripe width to thickness.

In the general case of a nanostripe with a spatially varying cross-section, the aspect ratio is coordinate-dependent, $\delta \equiv \delta(\xi)$. Consequently, the anisotropy parameters also become functions of the coordinate, $k_{\text{A}}(\xi)$ and $k_{\text{P}}(\xi)$.

We describe the magnetization direction using a unit vector $\boldsymbol{m}$ with the angular parametrization:

$$\boldsymbol{m} = \hat{\boldsymbol{x}}\cos\theta + \hat{\boldsymbol{y}}\sin\theta\cos\phi + \hat{\boldsymbol{z}}\sin\theta\sin\phi.$$

Using this parametrization, the normalized energy density can be written as:

$$\mathscr{E} = \mathcal{S}(\xi)\Big[\theta'^2 + \phi'^2\sin^2\theta + \sin^2\theta\left(k_{\text{A}} + k_{\text{P}}\sin^2\phi\right) - h\cos\theta\Big]. \tag{S6}$$

## SM-2. DOMAIN WALL DYNAMICS IN A NANOSTRIPE WITH VARYING CROSS SECTION

We describe the magnetization dynamics of this system using the Lagrange-Rayleigh formalism. The Lagrangian

* karakutsd@knu.ua
† k.yershov@ifw-dresden.de
‡ sheka@knu.ua

is given by:

$$\mathcal{L} = -\int_{-\infty}^{\infty} \mathcal{S}(\xi)\phi \sin\theta\dot{\theta}\mathrm{d}\xi - E, \tag{S7}$$

and the Rayleigh dissipative function is defined as:

$$\mathcal{F} = \frac{\alpha}{2}\int_{-\infty}^{\infty} \mathcal{S}(\xi)\left(\dot{\theta}^2 + \sin^2\theta\dot{\phi}^2\right)\mathrm{d}\xi, \tag{S8}$$

where $\alpha$ is the Gilbert damping coefficient. The dot denotes the derivative with respect to dimensionless time $\bar{t} = \omega_0 t$, where $\omega_0 = 4\pi\gamma_0 M_s$ determines the characteristic frequency of the system, and $\gamma_0$ is the gyromagnetic ratio.

To describe the domain wall (DW) dynamics in this system, we apply the collective variable approach as outlined in the main text. Substituting the ansatz (Eq. (1) of the main text) into the normalized energy density (S6) and integrating over the entire stripe yields:

$$E^{\mathrm{DW}} = h\int_{-\infty}^{\infty} \mathcal{S}(\xi)\tanh\frac{\xi - q}{\Delta}\mathrm{d}\xi + \left[\frac{1}{\Delta} + \Delta\left(k_{\mathrm{A}} + k_{\mathrm{P}}\sin^2\Phi\right)\right]\int_{-\infty}^{\infty} \mathcal{S}(\xi)\frac{\operatorname{sech}^2\frac{\xi-q}{\Delta}}{\Delta}\mathrm{d}\xi. \tag{S9}$$

Integration by parts of the first integral gives:

$$\int_{-\infty}^{\infty} h\mathcal{S}(\xi)\tanh\frac{\xi - q}{\Delta}\mathrm{d}\xi = -h\int_{-\infty}^{\infty} \mathcal{I}(\xi)\frac{\operatorname{sech}^2\frac{\xi-q}{\Delta}}{\Delta}\mathrm{d}\xi + \text{const}, \tag{S10}$$

where $\mathcal{I}(\xi)$ represents the derivative of the cross-sectional area profile. For a very narrow DW width and smooth $\mathcal{S}(\xi)$, we employ the following approximations:

$$\int_{-\infty}^{\infty} \mathcal{S}(\xi)\frac{\operatorname{sech}^2\frac{\xi-q}{\Delta}}{\Delta}\mathrm{d}\xi = 2\mathcal{S}(q), \quad \int_{-\infty}^{\infty} \mathcal{I}(\xi)\frac{\operatorname{sech}^2\frac{\xi-q}{\Delta}}{\Delta}\mathrm{d}\xi = 2\mathcal{I}(q), \tag{S11}$$

which leads to the effective DW energy (up to an additive constant), consistent with Eq. (3) of the main text. Although this approximation significantly simplifies the results and their interpretation, it is not strictly necessary, as Section SM-3 presents more precise equations.

By substituting the ansatz into the Lagrangian (S7),

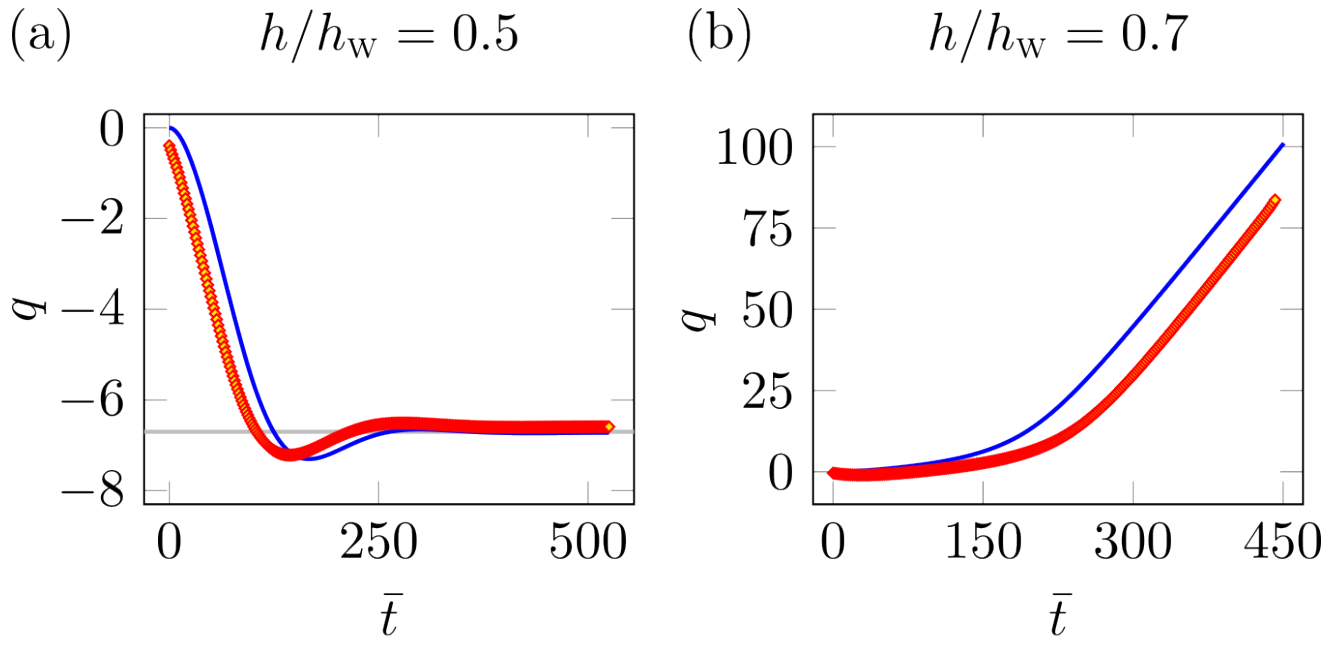


FIG. S1. DW motion in a wedge-shaped nanostripe under an applied magnetic field: (a) small field case ($h < h_d$); (b) large field case ($h > h_d$). The blue curves depict numerical solutions of the equations of motion (S25), while the symbols represent results from full-scale micromagnetic simulations (see Sec. SM-6). The gray line indicates the predicted pinning position (S26).

we obtain the effective Lagrangian of the domain wall:

$$\mathcal{L}^{\mathrm{DW}} = \Phi\dot{q}\int_{-\infty}^{\infty} \mathcal{S}(\xi)\frac{\operatorname{sech}^2\frac{\xi-q}{\Delta}}{\Delta}\mathrm{d}\xi - E^{\mathrm{DW}} + \Phi\dot{\Delta}\int_{-\infty}^{\infty} \mathcal{S}(\xi)\frac{\xi - q}{\Delta}\frac{\operatorname{sech}^2\frac{\xi-q}{\Delta}}{\Delta}\mathrm{d}\xi. \tag{S12}$$

Applying the same procedure to the dissipative function (S8), we obtain:

$$\mathcal{F}^{\mathrm{DW}} = \frac{\alpha\dot{\Delta}^2}{2\Delta}\int_{-\infty}^{\infty} \mathcal{S}(\xi)\left(\frac{\xi-q}{\Delta}\right)^2\frac{\operatorname{sech}^2\frac{\xi-q}{\Delta}}{\Delta}\mathrm{d}\xi + \frac{\alpha\dot{\Delta}\dot{q}}{\Delta}\int_{-\infty}^{\infty} \mathcal{S}(\xi)\left(\frac{\xi-q}{\Delta}\right)\frac{\operatorname{sech}^2\frac{\xi-q}{\Delta}}{\Delta}\mathrm{d}\xi + \frac{\alpha}{2\Delta}\left(\dot{q}^2 + \Delta^2\dot{\Phi}^2\right)\int_{-\infty}^{\infty} \mathcal{S}(\xi)\frac{\operatorname{sech}^2\frac{\xi-q}{\Delta}}{\Delta}\mathrm{d}\xi. \tag{S13}$$

For a very narrow DW width and smooth $\mathcal{S}(\xi)$, we use the approximations:

$$\int_{-\infty}^{\infty} \mathcal{S}(\xi)\left(\frac{\xi-q}{\Delta}\right)^2\frac{\operatorname{sech}^2\frac{\xi-q}{\Delta}}{\Delta}\mathrm{d}\xi = \frac{\pi^2}{6}\mathcal{S}(q), \tag{S14}$$

$$\int_{-\infty}^{\infty} \mathcal{S}(\xi)\left(\frac{\xi-q}{\Delta}\right)\frac{\operatorname{sech}^2\frac{\xi-q}{\Delta}}{\Delta}\mathrm{d}\xi = 0. \tag{S15}$$

Thus, the effective Lagrangian and dissipative function in terms of collective variables are:

$$\mathcal{L}^{\mathrm{DW}} = 2\mathcal{S}(q)\Phi\dot{q} - E^{\mathrm{DW}}, \tag{S16}$$

$$\mathcal{F}^{\mathrm{DW}} = \frac{\alpha}{\Delta}\mathcal{S}(q)\left[\dot{q}^2 + \left(\Delta\dot{\Phi}\right)^2 + c\dot{\Delta}^2\right], \tag{S17}$$

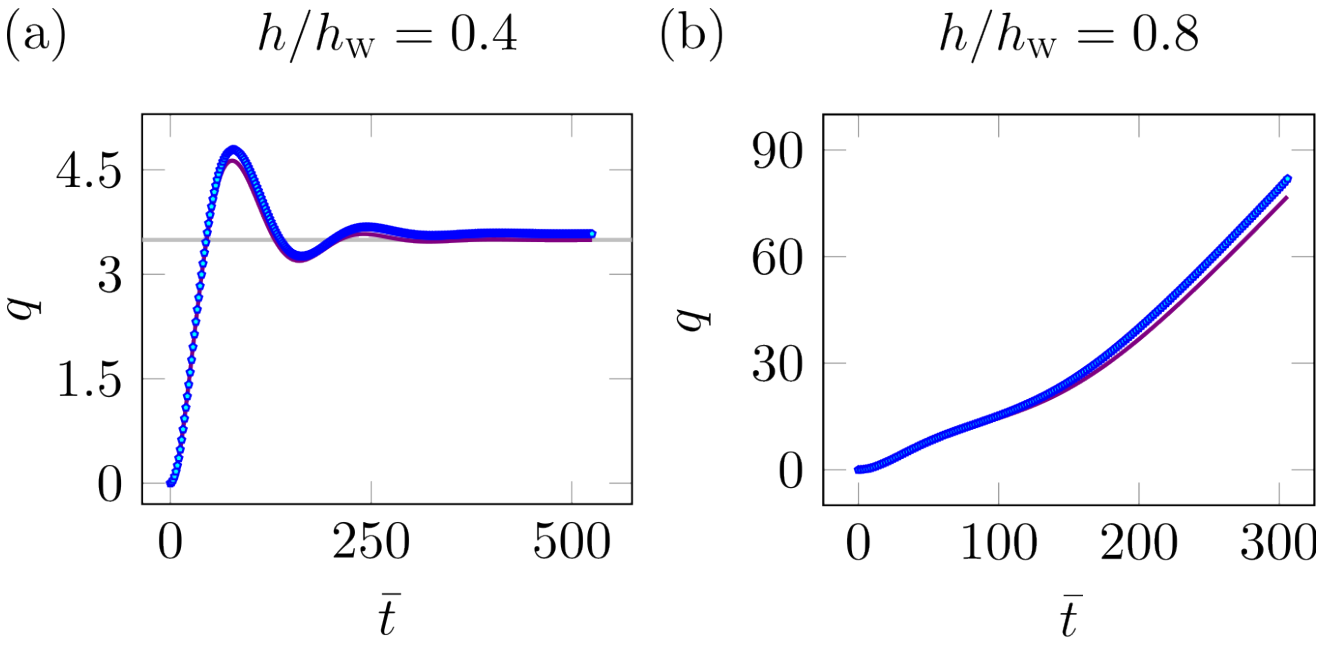


FIG. S2. DW motion in a nanostripe with a localized constriction under an applied magnetic field: (a) small field case ($h < h_d$); (b) large field case ($h > h_d$). The purple curves depict numerical solutions of the equations of motion (S25), while the symbols represent results from full-scale micromagnetic simulations (see Sec. SM-6). The gray line indicates the predicted pinning position (S26).

where $c = \pi^2/12$. Applying the Lagrange-Rayleigh formalism:

$$\frac{\delta \mathscr{L}}{\delta X_i} - \frac{\partial}{\partial \bar{t}} \frac{\delta \mathscr{L}}{\delta \dot{X}_i} = \frac{\delta \mathscr{F}}{\delta \dot{X}_i}, \qquad X_i = \{q, \Phi, \Delta\}, \tag{S18}$$

to Eqs. (S16) and (S17), we derive the equations of motion for the DW under the influence of an external magnetic field:

$$\begin{aligned} \alpha \frac{\dot{q}}{\Delta} + \dot{\Phi} &= \mathscr{F}(q), \\ \frac{\dot{q}}{\Delta} - \alpha \dot{\Phi} &= k_{\mathrm{P}} \sin 2\Phi, \\ \alpha c \dot{\Delta} &= \frac{\mathcal{S}'(q)}{\mathcal{S}(q)\Delta} - \Delta \left( k_{\mathrm{A}} + k_{\mathrm{P}} \sin^2 \Phi \right). \end{aligned} \tag{S19}$$

Note that the last equation describes the relaxation of $\Delta$ towards its instantaneous equilibrium value $\widetilde{\Delta}$. Linearization near this equilibrium shows that the characteristic relaxation time is $\tau = c\alpha\widetilde{\Delta}^2/2$. Since the damping parameter is typically very small ($\alpha \ll 1$), $\tau$ is also very small, implying that the DW width relaxes to its instantaneous equilibrium value rapidly. Consequently, we simplify the analysis by considering only the first two equations, assuming $\Delta(\bar{t}) \equiv \Delta\left[\Phi(\bar{t})\right] = 1/\sqrt{k_{\mathrm{A}} + k_{\mathrm{P}} \sin^2 \Phi}$ [4].

Furthermore, these equations can be refined by calculating the DW energy without the narrow DW approximation, as detailed in the next section.

## SM-3. NON-NARROW DOMAIN WALL IN A NANOSTRIPE WITH VARYING CROSS SECTION

We determine the DW energy in a wedge-shaped nanostripe without the narrow DW approximation. Although the magnetic energy density integral over $\xi \in (-\infty, \infty)$ is divergent in the general case, assuming the nanostripe is symmetric with respect to $\xi = 0$ allows for a well-defined evaluation.

We define the effective DW energy in a wedge-shaped nanostripe as:

$$E^{\mathrm{DW}} = \lim_{L \to +\infty} \int_{-L}^{L} \mathscr{E}(\xi) \mathrm{d}\xi, \tag{S20}$$

which allows Eq. (S9) to be rewritten in the form:

$$E^{\mathrm{DW}} = 2\text{ꝁ}_1 \left[ \frac{1}{\Delta} + \Delta \left( k_{\mathrm{A}} + k_{\mathrm{P}} \sin^2 \Phi \right) \right] - 2h\text{ꝁ}_2, \tag{S21}$$

where we introduce functions $\text{ꝁ}_i \equiv \text{ꝁ}_i(q)$ for $i = 1, 2, 3, 4$, defined as:

$$\begin{aligned} \text{ꝁ}_1 &= \lim_{L \to \infty} \left[ \frac{1}{2} \int_{-L}^{L} \mathcal{S}(x) \frac{\operatorname{sech}^2 \frac{x-q}{\Delta}}{\Delta} \mathrm{d}x \right], \\ \text{ꝁ}_2 &= \lim_{L \to \infty} \left[ \frac{1}{2} \int_{-L}^{L} \mathcal{S}(x) \tanh \frac{x-q}{\Delta} \mathrm{d}x \right], \\ \text{ꝁ}_3 &= \lim_{L \to \infty} \left[ \frac{1}{2} \int_{-L}^{L} \mathcal{S}(x) \frac{x-q}{\Delta} \frac{\operatorname{sech}^2 \frac{x-q}{\Delta}}{\Delta} \mathrm{d}x \right], \\ \text{ꝁ}_4 &= \lim_{L \to \infty} \left[ \frac{1}{2} \int_{-L}^{L} \mathcal{S}(x) \left( \frac{x-q}{\Delta} \right)^2 \frac{\operatorname{sech}^2 \frac{x-q}{\Delta}}{\Delta} \mathrm{d}x \right]. \end{aligned} \tag{S22}$$

Applying the ansatz (Eq. (1) of the main text) to the Lagrangian (S7) and dissipative function (S8), and integrating without the narrow DW approximation, yields the effective Lagrangian and dissipative function:

$$\begin{aligned} \mathscr{L}^{\mathrm{DW}} &= 2\Phi \left( \dot{q}\text{ꝁ}_1 + \dot{\Delta}\text{ꝁ}_3 \right) - E^{\mathrm{DW}}, \\ \mathscr{F}^{\mathrm{DW}} &= \frac{\alpha \text{ꝁ}_1}{\Delta} \left( \dot{q}^2 + \Delta^2 \dot{\Phi}^2 \right) + \frac{\alpha \dot{\Delta}}{\Delta} \left( 2\text{ꝁ}_3 \dot{q} + \dot{\Delta} \text{ꝁ}_4 \right). \end{aligned} \tag{S23}$$

Using the Lagrange-Rayleigh formalism (S18) on Eqs. (S23), we obtain the equations of motion:

$$\begin{aligned} &\frac{\alpha}{\Delta}\dot{q} + \dot{\Phi} + \left[ \frac{\alpha}{\Delta}\text{ꝁ}_3 + \Phi \left( \partial_\Delta \text{ꝁ}_1 - \partial_q \text{ꝁ}_3 \right) \right] \frac{\dot{\Delta}}{\text{ꝁ}_1} = \\ &h - \frac{\partial_q \text{ꝁ}_1}{\text{ꝁ}_1} \left[ \frac{1}{\Delta} + \Delta \left( k_{\mathrm{A}} + k_{\mathrm{P}} \sin^2 \Phi \right) \right], \\ &\frac{\dot{q}}{\Delta} - \alpha \dot{\Phi} + \frac{\text{ꝁ}_3}{\text{ꝁ}_1} \frac{\dot{\Delta}}{\Delta} = k_{\mathrm{P}} \sin 2\Phi, \\ &\frac{\alpha \dot{\Delta}}{\Delta} \text{ꝁ}_4 + \Phi \dot{q} \left( \partial_q \text{ꝁ}_3 - \partial_\Delta \text{ꝁ}_1 \right) + \text{ꝁ}_3 \left( \frac{\alpha \dot{q}}{\Delta} + \dot{\Phi} - h \right) = \\ &\frac{1}{\Delta^2} \left( \text{ꝁ}_1 - \Delta \partial_\Delta \text{ꝁ}_1 \right) - \left( \text{ꝁ}_1 + \Delta \partial_\Delta \text{ꝁ}_1 \right) \left( k_{\mathrm{A}} + k_{\mathrm{P}} \sin^2 \Phi \right). \end{aligned} \tag{S24}$$

In the general case, eliminating the third equation is not straightforward, as was possible in the previous section. However, we can extend the narrow DW

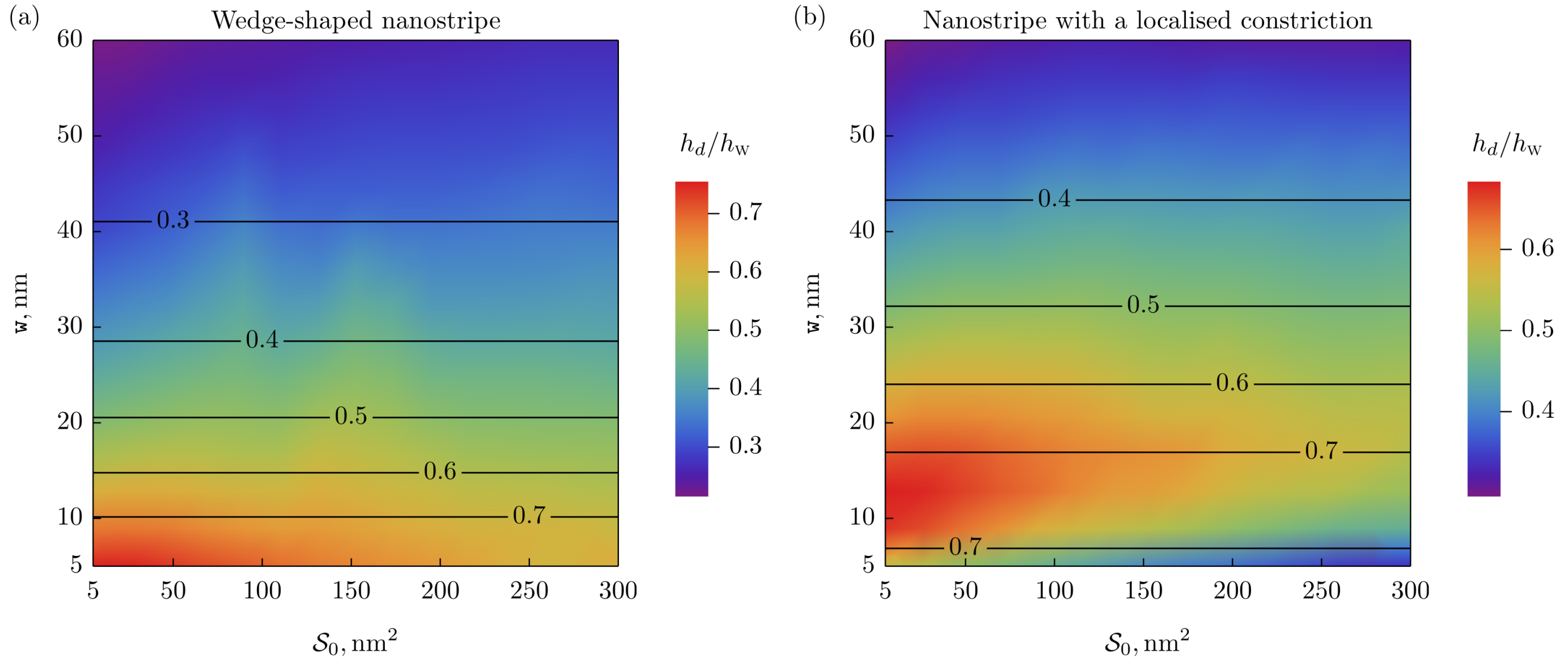


FIG. S3. Parameter dependence of the domain-wall depinning field and theoretical predictions. Panels (a) and (b) show two-dimensional parameter maps of DW depinning fields obtained from NMAG [6] micromagnetic simulations for a wedge-shaped nanostripe ($\mathfrak{d} = 0.05$, $\alpha = 0.07$) and a nanostripe with a localized constriction ($\mathfrak{d} = 0.25$, $\alpha = 0.15$), respectively. The color scale indicates the applied field required for DW depinning, normalized by $h_{\mathrm{W}} = \alpha k_{\mathrm{P}}$. The isolines demarcate theoretical predictions for the depinning field based on the non-narrow DW model (S27).

model by assuming that the DW width relaxes rapidly to $\Delta = 1/\sqrt{k_{\mathrm{A}}(q) + k_{\mathrm{P}}(q)\sin^2\Phi}$. Substituting this into the first two equations yields a simplified set:

$$\begin{aligned} &\frac{\alpha}{\Delta}\dot{q} + \dot{\Phi} = h - \frac{2}{\Delta}\frac{\partial_q \varkappa_1}{\varkappa_1}, \\ &\frac{\dot{q}}{\Delta} - \alpha\dot{\Phi} = k_{\mathrm{P}}\sin 2\Phi. \end{aligned} \tag{S25}$$

Numerical solutions of these equations for small and large external magnetic fields are displayed in Fig. S1.

The refined DW pinning conditions are obtained by setting $\dot{q} = 0$ and $\dot{\Phi} = 0$ in Eq. (S25):

$$\begin{aligned} \frac{2}{\Delta_\star}\partial_q \varkappa_1(q_p) &= h\varkappa_1(q_p), \\ \Phi_p &= \{0;\pi\}. \end{aligned} \tag{S26}$$

The refined depinning field is given by:

$$h_d = \frac{2}{\Delta_\star}\max_{q\in[q_0,+\infty)}\left[\frac{\partial_q \varkappa_1(q)}{\varkappa_1(q)}\right], \tag{S27}$$

where the DW width is $\Delta_\star(q) = 1/\sqrt{k_{\mathrm{A}}(q)}$. See Fig. S3 for details.

The equations of motion (S25) can be rewritten as:

$$\begin{aligned} &\frac{\alpha}{\Delta}\dot{q} + \dot{\Phi} = \mathscr{R}(q,\Phi), \\ &\frac{\dot{q}}{\Delta} - \alpha\dot{\Phi} = k_{\mathrm{P}}\sin 2\Phi, \end{aligned} \tag{S28}$$

where the effective field function $\mathscr{R}(q,\Phi)$ is defined as:

$$\mathscr{R}(q,\Phi) = -\frac{2}{\Delta}\frac{\partial_q \varkappa_1}{\varkappa_1} + h. \tag{S29}$$

We consider small deviations $\{\widetilde{q}, \widetilde{\Phi}\}$ of the DW position and phase from their equilibrium values. Expanding $\mathscr{R}$ and discarding nonlinear terms in $\widetilde{q}$ and $\widetilde{\Phi}$ yields linearized equations of motion. Using modified characteristic frequency components:

$$\begin{aligned} \Omega_g &= \frac{1}{1+\alpha^2}\left(\frac{2\partial_q^2 \varkappa_1}{\varkappa_1} - \frac{\Delta_p^2 h^2}{2}\right)\Bigg|_{\substack{q=q_p,\\ \Phi=\Phi_p}} \\ \Omega_{\mathrm{B}} &= \frac{2}{1+\alpha^2}k_{\mathrm{P}}, \end{aligned} \tag{S30}$$

we find that motion near the pinning position corresponds to damped oscillations with frequency $\Omega = \sqrt{\Omega_g\Omega_{\mathrm{B}} - \frac{\alpha^2(\Omega_g-\Omega_{\mathrm{B}})^2}{4}} \approx \sqrt{\Omega_g\Omega_{\mathrm{B}}}$ and decay rate $\eta = \frac{\alpha(\Omega_g+\Omega_{\mathrm{B}})}{2}$.

Analyzing Eq. (S25) near the pinning position $q_p$ in the absence of damping ($\alpha \to 0$), we assume the DW phase deviates by a small angle $\varphi$, such that $\Phi = \Phi_p + \varphi$. Substituting $v = 2k_{\mathrm{P}}\Delta_p\varphi$ into the effective DW energy (S21) yields:

$$E^{\mathrm{DW}} = E_{\mathrm{const}} + \mathcal{S}(q_p)\varkappa_1(q_p)\frac{v^2}{2k_{\mathrm{P}}\Delta_p}. \tag{S31}$$

Similar to Sec. I of the main text, this indicates that the DW possesses an inertial mass modified by cross-sectional variations:

$$m_{\mathrm{D}} = \varkappa_1(q_p)\frac{\mathcal{S}(q_p)}{k_{\mathrm{P}}\Delta_p}. \tag{S32}$$

The DW energy in a nanostripe with a localized constriction (up to an additive constant) is similar to Eq. (S21). Consequently, the equations derived in this section are also applicable to nanostripes with localized constrictions. However, as discussed in Sec. IV, these equations may describe significantly different behavior compared to the wedge-shaped case.

## SM-4. NARROW DOMAIN WALL IN A NANOSTRIPE WITH VARYING ASPECT RATIO

We describe how the magnetization dynamics of the nanostripe changes when the aspect ratio between the width and thickness is non-constant. In this case, the normalized energy density (S6) is modified, with the anisotropy coefficients becoming functions of the arclength coordinate, $k_{\mathrm{A}}(\xi)$ and $k_{\mathrm{P}}(\xi)$.

We derive the effective DW energy following the procedure outlined in Sec. SM-2. Substituting the ansatz (Eq. (1) of the main text) into the modified energy density (S6) and integrating over the entire stripe yields:

$$\begin{aligned} E^{\mathrm{DW}} &= h\int_{-\infty}^{\infty}\mathcal{S}(\xi)\tanh\frac{\xi-q}{\Delta}\mathrm{d}\xi \\ &+\frac{1}{\Delta}\int_{-\infty}^{\infty}\mathcal{S}(\xi)\frac{\operatorname{sech}^2\frac{\xi-q}{\Delta}}{\Delta}\mathrm{d}\xi \\ &+\Delta\int_{-\infty}^{\infty}\mathcal{S}(\xi)k_{\mathrm{A}}(\xi)\frac{\operatorname{sech}^2\frac{\xi-q}{\Delta}}{\Delta}\mathrm{d}\xi \\ &+\Delta\sin^2\Phi\int_{-\infty}^{\infty}\mathcal{S}(\xi)k_{\mathrm{P}}(\xi)\frac{\operatorname{sech}^2\frac{\xi-q}{\Delta}}{\Delta}\mathrm{d}\xi. \end{aligned} \tag{S33}$$

The integration by parts of the first integral is given in Eq. (S10). For a very narrow DW width, satisfying $\Delta \ll \mathcal{S}/\mathcal{S}' + k_{\mathrm{A}}/k_{\mathrm{A}}'$ and $\Delta \ll \mathcal{S}/\mathcal{S}' + k_{\mathrm{P}}/k_{\mathrm{P}}'$, we use the approximations:

$$\int_{-\infty}^{\infty}\mathcal{S}(\xi)k_{\mathrm{A}}(\xi)\frac{\operatorname{sech}^2\frac{\xi-q}{\Delta}}{\Delta}\mathrm{d}\xi = 2\mathcal{S}(q)k_{\mathrm{A}}(q), \tag{S34}$$

$$\int_{-\infty}^{\infty}\mathcal{S}(\xi)k_{\mathrm{P}}(\xi)\frac{\operatorname{sech}^2\frac{\xi-q}{\Delta}}{\Delta}\mathrm{d}\xi = 2\mathcal{S}(q)k_{\mathrm{P}}(q). \tag{S35}$$

Combining these approximations with Eq. (S11) allows us to evaluate Eq. (S33), resulting in the effective DW energy [Eq. (14) of the main text]. Although the effective DW energy has changed, the Lagrangian and dissipative function retain the same forms as in Eqs. (S7) and (S8). Application of the formalism (S18) yields the modified equations of motion for a DW in a nanostripe with a non-constant aspect ratio [Eq. (15) of the main text].

These equations can be further refined by calculating the DW energy without the narrow DW approximation, as detailed in the next section.

## SM-5. NON-NARROW DOMAIN WALL IN A NANOSTRIPE WITH VARYING ASPECT RATIO

We determine the DW energy in a nanostripe with varying aspect ratio without the narrow DW approximation. Introducing new functions $\mathcal{g}_i \equiv \mathcal{g}_i(q)$ for $i = \mathrm{A}, \mathrm{P}$, defined as:

$$\begin{aligned} \mathcal{g}_{\mathrm{A}} &= \lim_{L\to\infty}\left[\frac{1}{2}\int_{-L}^{L}\mathcal{S}(x)k_{\mathrm{A}}(x)\frac{\operatorname{sech}^2\frac{x-q}{\Delta}}{\Delta}\mathrm{d}x\right], \\ \mathcal{g}_{\mathrm{P}} &= \lim_{L\to\infty}\left[\frac{1}{2}\int_{-L}^{L}\mathcal{S}(x)k_{\mathrm{P}}(x)\frac{\operatorname{sech}^2\frac{x-q}{\Delta}}{\Delta}\mathrm{d}x\right], \end{aligned} \tag{S36}$$

we obtain the effective DW energy in the form:

$$E^{\mathrm{DW}} = 2\left[\frac{\mathcal{k}_1}{\Delta} + \Delta\left(\mathcal{g}_{\mathrm{A}} + \mathcal{g}_{\mathrm{P}}\sin^2\Phi\right) - h\mathcal{k}_2\right]. \tag{S37}$$

The Lagrangian and dissipative function retain the forms given in Eq. (S23), albeit with the new effective DW energy. Application of the formalism (S18) yields the full three-parameter equations of motion for a non-constant aspect ratio without the narrow DW approximation:

$$\begin{aligned} &\frac{\alpha}{\Delta}\dot{q} + \dot{\Phi} + \left[\frac{\alpha}{\Delta}\mathcal{k}_3 + \Phi\left(\partial_\Delta\mathcal{k}_1 - \partial_q\mathcal{k}_3\right)\right]\frac{\dot{\Delta}}{\mathcal{k}_1} = \\ &h - \frac{1}{\Delta}\frac{\partial_q\mathcal{k}_1}{\mathcal{k}_1} - \frac{\Delta}{\mathcal{k}_1}\left(\partial_q\mathcal{g}_{\mathrm{A}} + \partial_q\mathcal{g}_{\mathrm{P}}\sin^2\Phi\right), \\ &\frac{\dot{q}}{\Delta} - \alpha\dot{\Phi} + \frac{\mathcal{k}_3}{\mathcal{k}_1}\frac{\dot{\Delta}}{\Delta} = \frac{\mathcal{g}_{\mathrm{P}}}{\mathcal{k}_1}\sin 2\Phi, \\ &\frac{\alpha\dot{\Delta}}{\Delta}\mathcal{k}_4 + \Phi\dot{q}\left(\partial_q\mathcal{k}_3 - \partial_\Delta\mathcal{k}_1\right) + \mathcal{k}_3\left(\frac{\alpha\dot{q}}{\Delta} + \dot{\Phi} - h\right) = \\ &\frac{1}{\Delta^2}\left(\mathcal{k}_1 - \Delta\partial_\Delta\mathcal{k}_1\right) - \left(\mathcal{g}_{\mathrm{A}} + \mathcal{g}_{\mathrm{P}}\sin^2\Phi\right) \\ &-\Delta\left(\partial_\Delta\mathcal{g}_{\mathrm{A}} + \partial_\Delta\mathcal{g}_{\mathrm{P}}\sin^2\Phi\right). \end{aligned} \tag{S38}$$

## SM-6. NUMERICAL SIMULATIONS

To verify our theoretical predictions, we performed full-scale micromagnetic simulations of the Landau-Lifshitz-Gilbert equation using the NMAG code [6]. We considered only exchange and magnetostatic interactions, neglecting thermal effects.

The material parameters were chosen to correspond to Permalloy: the exchange constant $A = 2.6\times10^{-6}$ erg/cm (26 pJ/m) and the saturation magnetization $M_s = 800$ G (800 kA/m). These parameters result in an exchange

length $\ell \approx 5.7$ nm and a characteristic frequency $\frac{\omega_0}{2\pi} \approx 28.14$ GHz.

Micromagnetic simulations were conducted for two types of nanostripes with a constant aspect ratio $\delta = w/d = 3$: a nanostripe with a localized constriction and a wedge-shaped nanostripe. Additionally, we simulated a nanostripe with a localized constriction created solely by varying the thickness, with a base aspect ratio $\delta_0 = w_0/d_0 = 3$. In all cases, nanostripes of length $L = 2000$ nm ($s \in [-1000, 1000]$) were modeled using irregular tetrahedral meshes with a characteristic cell size of 1 nm.

The nanostripe geometry was defined by specifying the cross-sectional area function $\mathcal{S}(x)$, which determines the corresponding width $w$ and thickness $d$ at each point, as described in Secs. IV and V of the main text. A wedge-shaped nanostripe was modeled using Eq. (12) of the main text. A nanostripe with a localized constriction was described by Eq. (13) for the constant aspect ratio case and Eq. (22) for the solely varying thickness case.

The simulations proceeded in two stages. First, DW relaxation was performed in the overdamped regime ($\alpha = 0.5$) without an external field, with the DW initially located at $q_0 = 0$. In the second stage, DW dynamics in the presence of an external field were simulated with a small damping coefficient $\alpha = 0.07$.

For all cases, the simulations covered a series of applied magnetic fields defined as fractions of the Walker field $h_{\mathrm{W}} = \alpha k_{\mathrm{P}}$ [for the solely varying thickness case, $h_{\mathrm{W}} = \alpha k_{\mathrm{P}}(+\infty)$], in the range $h \in [0,\ 0.9\, h_{\mathrm{W}}]$ with a step of $0.1\, h_{\mathrm{W}}$.

The simulation duration was 5 ns, during which the tangential component of the magnetization was recorded. The obtained data were approximated by the $q$-$\Phi$ model to extract the DW position $q$ at each time instant. The DW positions at the end of the simulations for the pinning cases are shown in Fig. 2(a), Fig. 3(a), and Fig. 4(a).

The asymptotic DW velocity was obtained by fitting the linear portion of the DW motion with a straight line; the slope of this line provides the DW velocity. The resulting velocities, converted to dimensionless units, are presented in Fig. 2(c), Fig. 3(c), and Fig. 4(c).

The frequency of DW oscillations near the pinning position, $\Omega$, was determined by approximating the DW motion with a damped periodic function:

$$[a \cos(\Omega t) + b \sin(\Omega t)]\, e^{-\eta t} + c, \tag{S39}$$

where the angular frequencies, converted to dimensionless units, are shown in Fig. 2(b), Fig. 3(b), and Fig. 4(b).

Additionally, a separate series of simulations was performed to evaluate how the model's accuracy varies with increasing cross-sectional area and variation length. Simulations were conducted over a wide range of cross-sectional areas and variation lengths, with $\mathcal{S}_0 \in [5; 300]$ nm$^2$ and variation width $w \in [5; 60]$ nm.

Using our theoretical predictions as a starting point, we identified the field at which the transition to a depinned state occurs with a precision of $\Delta h = 0.01\, h_{\mathrm{W}}$. Note that the parameters for these simulations differed slightly from the previous set: the expansion magnitude for the wedge-shaped nanostripe was $\mathfrak{s} = 0.05$, and the damping coefficient for the second stage of the nanostripe with a localized constriction was $\alpha = 0.15$. Relative errors ($\epsilon = |h_{\text{theory}} - h_{\text{NMAG}}|/h_{\text{NMAG}}$) between the theoretical predictions and simulation results were evaluated, and the narrow-DW applicability cutoff is shown in Fig. 5 of the main text.


\clearpage
\end{document}